\documentclass[oldversion]{aa}  

\usepackage{graphicx}
\usepackage{txfonts}
\usepackage{float}
\usepackage[latin1]{inputenc}
\usepackage{supertabular}
\usepackage{color}
\usepackage{natbib}
\bibpunct{(}{)}{;}{a}{}{,}

\newcommand{\rmh}{^{\mathrm{h}}}
\newcommand{\rmm}{^{\mathrm{m}}}

\begin{document}

\title{The Hamburg/ESO R-process Enhanced Star survey (HERES) \thanks{Based on
    observations collected at the European Southern Observatory, Paranal,
    Chile (Proposal Number 170.D-0010).}  }

\subtitle{IV. Detailed abundance analysis and age dating of the strongly r-process enhanced 
          stars CS~29491$-$069 and HE~1219$-$0312.
         }

\author{
  W. Hayek\inst{1,2,3} \and
  U. Wiesendahl\inst{1} \and
  N. Christlieb\inst{4} \and
  K. Eriksson\inst{5} \and
  A.J. Korn\inst{5} \and
  P.S. Barklem\inst{5} \and
  V. Hill\inst{6} \and
  T.C. Beers\inst{7} \and
  K. Farouqi\inst{8} \and
  B. Pfeiffer\inst{9} \and
  K.-L. Kratz\inst{9}
}

\offprints{W. Hayek,\\ \email{hayek@mpa-garching.mpg.de}}
\institute{
Hamburger Sternwarte, Universit\"at Hamburg, Gojenbergsweg 112,
     D-21029 Hamburg, Germany 
\and Research School of Astronomy and Astrophysics, Mt. Stromlo Observatory,
     Cotter Rd., Weston Creek, ACT 2611, Australia 
\and Max Planck Institut f\"ur Astrophysik, Karl-Schwarzschild-Str. 1, D-85741 Garching, Germany
\and Zentrum f\"ur Astronomie der Universit\"at Heidelberg, Landessternwarte, K\"onigstuhl 12, D-69117 Heidelberg, Germany 
\and Department of Physics and Astronomy, Uppsala University, Box 515, 
     75120 Uppsala, Sweden 
\and Cassiop\'ee, Observatoire de la C\^{o}te d'Azur, CNRS, Universit\'e de Nice Sophia-Antipolis,
   Bd. de l'Observatoire, 06300 Nice, France
\and Department of Physics and Astronomy, CSCE: Center for the Study of Cosmic Evolution,
    and JINA: Joint Institute for Nuclear Astrophysics, Michigan State University, East
    Lansing, MI 48824, USA 
\and Department of Astrophysics and Astronomy, University of Chicago, Chicago IL 60637, USA
\and Max-Planck-Institut f\"ur Chemie (Otto-Hahn-Institut), Joh.-J. Becherweg 27, D-55128 Mainz, Germany 
}

\date{Received 9 October 2008 / Accepted 12 June 2009}

\abstract{    
We report on a detailed abundance analysis of two strongly r-process enhanced, very metal-poor stars newly discovered in the HERES project, \object{CS~29491$-$069} ($\mathrm{[Fe/H]}=-2.51$, $\mathrm{[r/Fe]}=+1.1$) and \object{HE~1219$-$0312} ($\mathrm{[Fe/H]}=-2.96$, $\mathrm{[r/Fe]}=+1.5$). The analysis is based on high-quality VLT/UVES spectra and MARCS model atmospheres. We detect lines of 15 heavy elements in the spectrum of \object{CS~29491$-$069}, and 18 in \object{HE~1219$-$0312}; in both cases including the \ion{Th}{II} 4019\,{\AA} line. The heavy-element abundance patterns of these two stars are mostly well-matched to scaled solar residual abundances not formed by the s-process. We also compare the observed pattern with recent high-entropy wind (HEW) calculations, which assume core-collapse supernovae of massive stars as the astrophysical environment for the r-process, and find good agreement for most lanthanides. The abundance ratios of the lighter elements strontium, yttrium, and zirconium, which are presumably not formed by the main r-process, are reproduced well by the model. Radioactive dating for \object{CS~29491$-$069} with the observed thorium and rare-earth element abundance pairs results in an average age of $9.5$\,Gyr, when based on solar r-process residuals, and $17.6$\,Gyr, when using HEW model predictions. Chronometry seems to fail in the case of \object{HE~1219$-$0312}, resulting in a negative age due to its high thorium abundance. \object{HE~1219$-$0312} could therefore exhibit an overabundance of the heaviest elements, which is sometimes called an ``actinide boost''.
}
 
\maketitle

\section{Introduction}\label{sec:intro}

The ages of the oldest stars in the Galaxy provide an important constraint for the time of the onset of stellar nucleosynthesis, with further implications for galaxy formation and evolution. Among the various methods used for age determinations of old stars, nucleochronometry based on long-lived radioactive isotopes, e.g. $^{232}$Th (half-life $\tau_{1/2}=14.05$\,Gyr) or $^{238}$U ($\tau_{1/2} = 4.468$\,Gyr), has attracted a lot of attention during the last decade. Radioactive decay ages can be derived by comparing the observed abundances of these elements, relative to a stable r-process element, to the production ratio expected from theoretical r-process yields.

The interest in this age determination method was increased by discoveries of very metal-poor stars that are strongly enhanced in r-process elements (i.e., stars having $\mbox{[Eu/Fe]} > +1.0$ and $\mbox{[Ba/Eu]} < 0$; hereafter r-II stars, following \citealt{Beers/Christlieb:2005}), for which the measurement of Th and Eu abundances was possible, and whose stellar matter presumably has experienced only a single nucleosynthesis event. The Th/Eu chronometer has been used for determining the age of, e.g., the progenitor of the r-II star \object{CS~22892$-$052} \citep{Snedenetal:1996,Cowanetal:1997,Cowanetal:1999,Snedenetal:2003}. Uranium was first detected in a very metal-poor star, \object{CS~31082$-$001}, by \citet{Cayreletal:2001}. \citet{Cowanetal:2002} tentatively detected U in \object{BD\,$+17^{\circ}3248$}, and recently \citet{Frebeletal:2007b} have clearly seen U in \object{HE~1523$-$0901}.  However, it is observationally very challenging to measure the uranium abundance of metal-poor stars, because even in cool, strongly r-process-enhanced, metal-poor giants the strongest U line observable with ground-based telescopes, at $\lambda = 3859.571$\,{\AA}, has an equivalent width of only a few m{\AA}. Therefore, very high signal-to-noise ratio ($S/N$) spectra are required. Furthermore, the line is detectable only in metal-poor stars without overabundances of carbon or nitrogen, since it is blended with a molecular CN feature. As a result, a U abundance can be measured only in about 1 out of $10^6$ halo stars, while the fraction of halo stars for which abundances of Th and Eu (or similar r-process elements) can be determined is about one order of magnitude larger.

However, a significant complication is that the Th/Eu chronometer seems to fail in some r-process enhanced metal-poor stars, resulting in negative age estimates. For example, \citet{Hilletal:2002} report $\log\left(\mbox{Th/Eu}\right) = -0.22$ for \object{CS~31082$-$001}, and \citet{Hondaetal:2004b} derive $\log\left(\mbox{Th/Eu}\right) = -0.10$ for \object{CS~30306$-$132}; the abundances of the heaviest r-process elements (third-peak-elements, actinides) seem enhanced in these stars with respect to the lanthanide Eu. By comparison, \citet{Snedenetal:2003} measure $\log\left(\mbox{Th/Eu}\right)=-0.62$ in \object{CS~22892$-$052}, and, using a production ratio of $\log\left(\mbox{Th/Eu}\right)_0 = -0.35$, derive an age of $12.8\pm3$\,Gyr. These results clearly cast doubts on the reliability of the Th/Eu chronometer pair; the observed relative abundances of intermediate and some heavy r-process elements diverge. For stars where uranium cannot be detected, third-peak elements such as osmium (if available) seem better partners for age determinations, as theoretical predictions of their corresponding r-process yields are more robust than those for pairs of heavy and intermediate mass elements. However, the dominant transitions of third-peak elements, which are detected as neutral species, lie in the UV. They are therefore hard to employ for reliable abundance analyses. Moreover, abundance ratios with rare-earth elements are more sensitive to uncertainties in the model atmosphere, since these are mostly measured using ionized species \citep[see also][]{Kratzetal:2007}.

Different astrophysical sites for the r-process have been suggested in the past, including core-collapse supernovae of massive stars, neutron-star mergers and more exotic candidates, but none of these scenarios have so far been proven. The para\-meters of r-process models therefore had to be defined site-independently; they were nevertheless able to reproduce observed abundance patterns of heavy neutron-capture elements in the Sun and metal-poor stars (see, e.g., papers by \cite{Kratzetal:1993}, \cite{Pfeifferetal:2001}, \citet{Wanajoetal:2002} and \citet{Kratzetal:2007}). Recent studies showed that lighter elements such as strontium, yttrium and zirconium exhibit more complex behavior, which cannot be explained in a simple r-process scenario. We compare our abundance measurements to the nucleosynthetic yields of a new generation of high-entropy wind (HEW) models which include additional charged-particle processes \citep{Farouqi:2005,Farouqietal:2005,Farouqietal:2008a,Farouqietal:2008b}.

This paper continues our series on the Hamburg/ESO R-process-Enhanced Star survey (HERES). A detailed description of the project and its aims can be found in \citet[][ hereafter Paper~I]{HERESpaperI}; methods of automated abundance analysis of high-resolution ``snapshot'' spectra have been described in \citet[][ hereafter Paper~II]{HERESpaperII}. In this paper we report on detailed abundance analyses of the r-II stars \object{CS~29491$-$069} and \object{HE~1219$-$0312} (Sect.~\ref{sec:AbundanceAnalysis}) based on high-quality VLT/UVES spectra (for details see Sect.~\ref{sec:ObservationReduction}) and MARCS model atmospheres. Our results are presented in Sect.~\ref{sec:Results} and discussed in Sect.~\ref{sec:DiscussionConclusions}.

\section{Observations and data reduction}\label{sec:ObservationReduction}

\subsection{Observations}\label{observation}

\begin{table}[t]
 \centering
 \caption{Coordinates and photometry of \object{HE~1219$-$0312} and 
   \object{CS~29491$-$069}.}
 \label{tab:CoordsPhotometry}
  \begin{tabular}{l r r}
   \hline\hline
   Quantity & \object{HE~1219$-$0312}        & \object{CS~29491$-$069}           \\\hline
   R.A.(2000.0) & $12\rmh 21\rmm 34\fs1$         & $22\rmh 31\rmm 02\fs1$         \\
   dec.(2000.0) & $-03\degr 28\arcmin 40\arcsec$ & $-32\degr 38\arcmin 36\arcsec$ \\
   $V$ [mag]    & $15.940 \pm 0.007$             & $13.075 \pm 0.002$             \\
   $B-V$        & $0.641  \pm 0.027$             & $0.600  \pm 0.004$             \\
   $V-R$        & $0.455  \pm 0.011$             & $0.421  \pm 0.003$             \\
   $V-I$        & $0.897  \pm 0.009$             & $0.900  \pm 0.004$             \\
   \hline
  \end{tabular}
\end{table}

\begin{figure}[b]
  \centering
  \resizebox{\hsize}{!}{\includegraphics{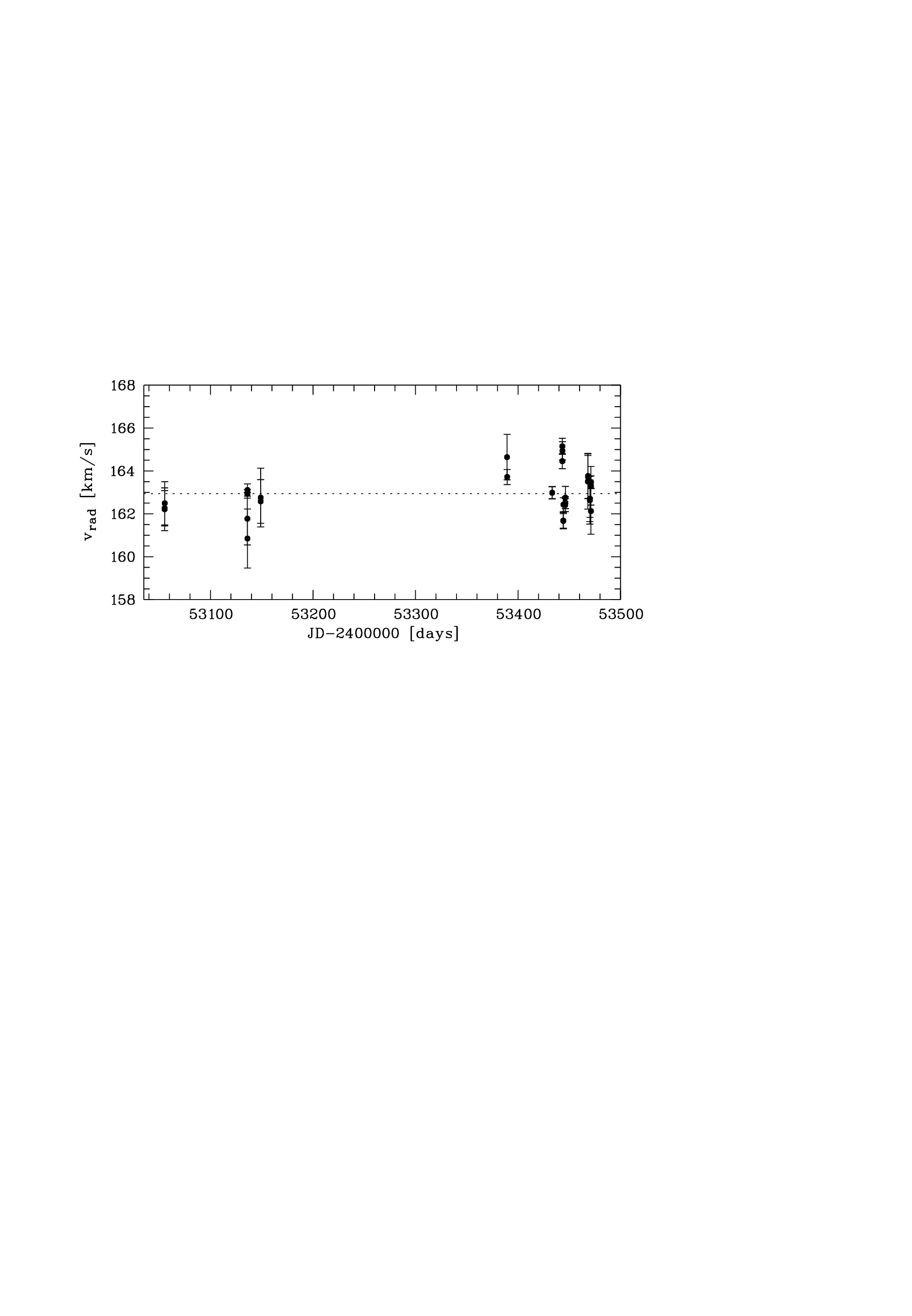}}
  \caption{\label{fig:HE1219_vrad_MJD} Barycentric radial velocity
    measurements of \object{HE~1219$-$0312}, spanning a period of $\sim 14$
    months. The error bars refer to the $1\sigma$ standard deviation of
    15 measurements in every B346 spectrum and 8 measurements in every B437 spectrum.} 
\end{figure}

\begin{table*}[htbp]
 \centering
 \caption{VLT/UVES observations of \object{HE~1219$-$0312} and \object{CS~29491$-$069}.}
 \label{tab:Observation}
  \begin{tabular}{l c c c c c c c}
   \hline\hline
   UVES & $\lambda$-range [nm] & \multicolumn{2}{c}{Total integration time [h]} & \multicolumn{2}{c}{Resolving power R} & \multicolumn{2}{c}{S/N ratio$^2$} \\
   Setup & Nominal$^1$ & \object{HE~1219$-$0312} & \object{CS~29491$-$069} & \object{HE~1219$-$0312} & \object{CS~29491$-$069} & \object{HE~1219$-$0312} &  \object{CS~29491$-$069} \\
   \hline
   BLUE346    &      $303-388$      &   16   & --    &  71,100     & --      &   24      & --   \\
   BLUE390    & $326-445$  & --   & 1   & --    & 57,500  & --      & 40   \\
   BLUE437    & $373-499$  &  15    & 1   &   71,400    & 57,600  &    50     & 70   \\[0.1cm]
   RED580$^3$ &            &      &       &       &         &         &      \\
   lower      & $476-580$  &   16   & 1   &   71,300    & 54,600  &    87     & 70   \\
   upper      & $582-684$  &         16           & 1   &    65,500   & 52,300  &    111     & 110  \\[0.1cm]
   RED860$^3$ &            &                    &       &       &         &         &      \\
   lower      & $660-854$  &          15          & 1   &   72,000    & 54,900  &     99    & 140  \\
   upper      & $865-1060$ &          15          & 1   &    66,700   & 52,100  &    50     & 40   \\
   \hline
   \multicolumn{8}{l}{1: from the UVES manual; 2: per pixel, at 3450\,{\AA}, 3700\,{\AA}, \ion{Th}{II} 4019\,{\AA}, H$\beta$, H$\alpha$, 7500\,{\AA} or 9500\,{\AA}; 3: two-CCD mosaic in red arm}\\                    
  \end{tabular}
\end{table*}

\object{CS~29491$-$069} and \object{HE~1219$-$0312} were identified as metal-poor stars in the HK survey of Beers et al. \citep{BPSI,BPSII} and the stellar part of the Hamburg/ESO Survey (HES; \citealt{Christliebetal:2008}), respectively. The coordinates and photometric data for both stars can be found in Table~\ref{tab:CoordsPhotometry}. The photometry was taken from \citet{Beersetal:2007}.

Since moderate-resolution (i.e., $\Delta\lambda \sim 2$\,{\AA}) follow-up observations indicated that both stars are cool giants having ${\rm [Fe/H]}\sim -3.0$, they were included in the target list of the HERES project. ``Snapshot'' high-resolution spectra (i.e., spectra having $S/N \sim 50$ per pixel at $4100$\,{\AA} and $R\sim 20,000$) revealed that both stars exhibit strong overabundances of the r-process elements ($\mathrm{[Eu/Fe]_{CS29491}}=+1.06$ and $\mathrm{[Eu/Fe]_{HE1219}}=+1.41$; see Paper~II); that is, both of them are r-II stars.

Spectra of higher $S/N$ and higher resolving power were obtained with VLT/UVES in Service Mode in 2004 and 2005. UVES was used in dichroic mode in various settings in order to achieve complete coverage of the optical wavelength range. Slit widths of $0.8''$ and $0.6''$ were chosen for \object{CS~29491$-$069} and \object{HE~1219$-$0312}, respectively, with the aim of reaching resolving powers of $R\sim 60,000$ and $R\sim 70,000$, respectively. Details of the applied settings, total integration times, and $S/N$ of the co-added spectra are listed in Table~\ref{tab:Observation}.

\subsection{Data reduction}\label{reduction}

We used pipeline-reduced UVES spectra for our analysis, as provided by the ESO Data Management Division. For comparison purposes, the spectra of \object{CS~29491$-$069} were also reduced with the \texttt{REDUCE} package of \citet{Piskunov:2002}.

The observations of \object{CS~29491$-$069} spanned only $\sim1.5$ months, and the barycentric radial velocities, measured in the individual spectra by fitting Gaussian profiles to 15 moderately strong and unblended lines throughout the spectral range, are consistent with each other to within the measurement uncertainties: $v_\mathrm{rad,bary}=-377.9\pm 1.0$\,km/s (MJD 53281.002) and $v_\mathrm{rad,bary}=-376.3\pm 1.0$\,km/s (MJD 53324.069).

\object{HE~1219$-$0312} was observed over a period of $\sim 14$ months. The measured barycentric radial velocities are shown in Fig.~\ref{fig:HE1219_vrad_MJD}. As in \object{CS~29491$-$069}, no significant radial velocity changes are seen.

Since 31 observing blocks were executed for \object{HE~1219$-$0312}, the individual spectra had to be co-added. After shifting all spectra to the rest frame, an iterative procedure was applied which identified all pixels in the individual spectra which were affected by cosmic ray hits, CCD defects, or other artifacts and not yet removed during the data reduction. These pixels were flagged and ignored in the final iteration of the co-addition. The $S/N$ of the resulting spectra is listed in Table~\ref{tab:Observation}.

\section{Abundance analysis}\label{sec:AbundanceAnalysis}

\subsection{Stellar parameters}\label{sec:stellarParameters}

\begin{table*}[htbp]
 \centering
 \caption{Adopted stellar parameters of \object{CS~29491$-$069} and   
          \object{HE~1219$-$0312} compared to the values determined
          by \citet{HERESpaperII}.} 
 \label{tab:StellarParameters}
 \begin{tabular}{lccccc}\hline\hline
   & \multicolumn{2}{c}{\object{CS~29491$-$069}} & \multicolumn{2}{c}{\object{HE~1219$-$0312}}\\
  Parameter & This work       & HERES           & This work      & HERES          \\\hline
   $ T_\mathrm{eff}$ [K] & $5300\pm100$    & $5103\pm100$    & $5060\pm100$   & $5140\pm100$   \\
   $ \log g$ (cgs)             & $2.8\pm0.2$        & $2.5\pm0.3$          & $2.3\pm0.2$    & $2.4\pm0.4$    \\
   $ [$Fe/H$]$                  & $-2.51\pm0.16$  & $-2.76\pm0.13$    & $-2.96\pm0.14$ & $-2.80\pm0.12$ \\
   $ \xi$ [km/s]                  & $1.6\pm0.2$        & $1.5\pm0.2$         & $1.6\pm0.1$    & $1.5\pm0.2$    \\\hline
  \end{tabular}
\end{table*}

\begin{figure}[b]
  \centering
  \resizebox{\hsize}{!}{\includegraphics{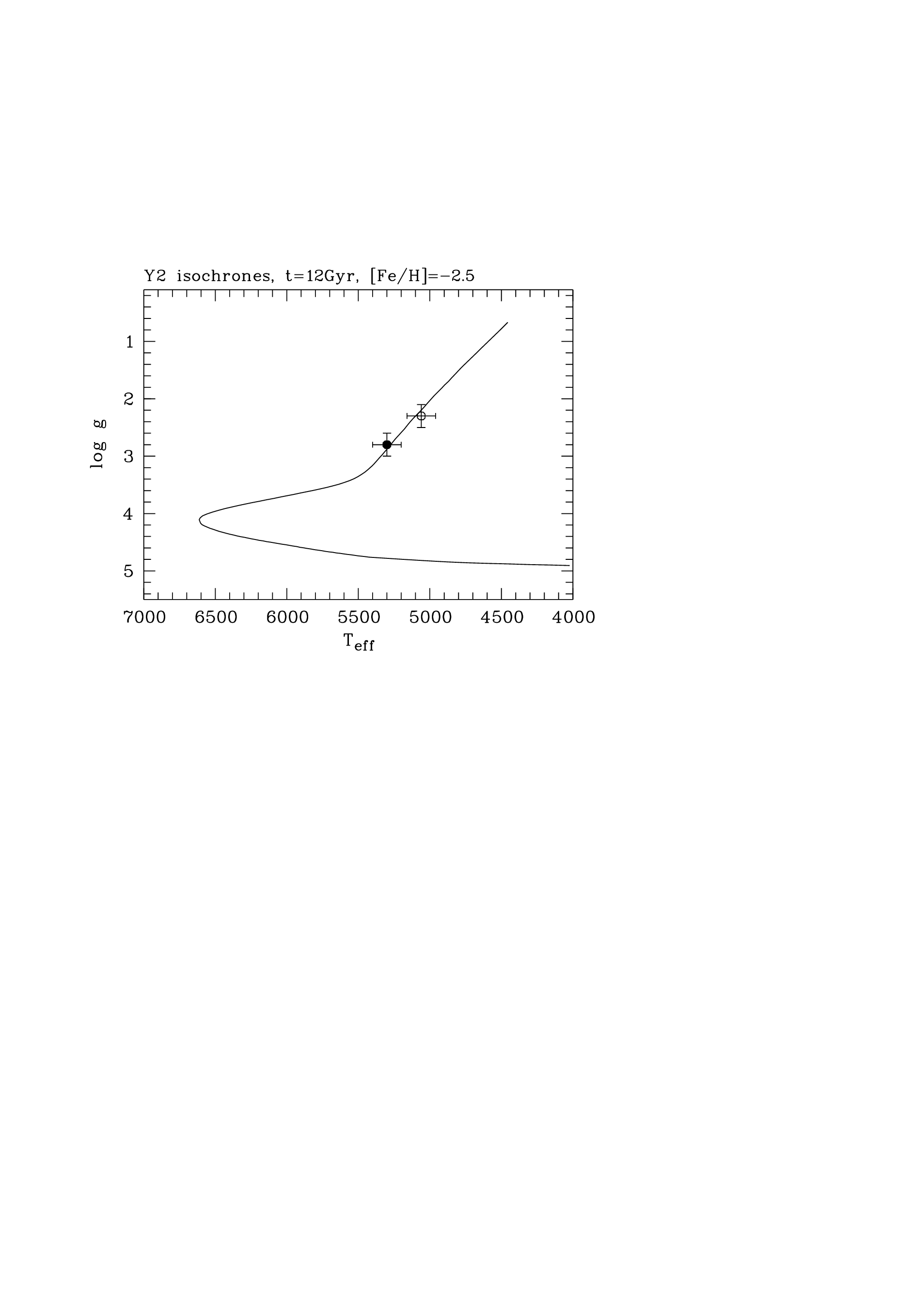}}
 \caption{Comparison of the adopted stellar parameters of \object{CS~29491$-$069} 
    (filled circle) and \object{HE~1219$-$0312} (open circle) with a 
    12\,Gyr, $\mathrm{[Fe/H]}=-2.5$ isochrone \citep{Yi:2001,Kim:2002}.}
  \label{fig:Isochrones}
\end{figure}

The stellar atmospheric parameters of \object{CS~29491$-$069} and \object{HE~1219$-$0312}, the effective temperature $T_{\rm eff}$, the surface gravity $\log g$, and the metallicity [Fe/H], were determined using the fully reduced spectra (single order \texttt{REDUCE} spectra in the case of \object{CS~29491$-$069}). A.J.K. and P.S.B. performed independent measurements with different techniques.

A.J.K. determined all three parameters following an iterative approach, thereby accounting for their interdependencies. Using the parameters from the previous ``snapshot'' spectroscopy and photometry (marked as HERES in Table~\ref{tab:StellarParameters}) as starting values and our new high-resolution spectra, the surface gravity $\log g$ was measured through the ionization equilibrium of \ion{Fe}{I} and \ion{Fe}{II}. Line formation was calculated with a 1D \texttt{MAFAGS} model atmosphere and in NLTE (non-local thermodynamic equilibrium), accounting for the over-ionization effect of \ion{Fe}{I}. The model iron atom includes 236 terms of \ion{Fe}{I} and 267 terms of \ion{Fe}{II}, as well as an empirically calibrated approximation
for inelastic hydrogen collisions. The microturbulence parameter $\xi$ was adjusted at the same time by requiring that the line abundances be independent of the absorption strengths. The metallicity [Fe/H] is a direct outcome of this procedure. 

As an example, the NLTE metallicity of \object{CS~29491$-$069} is [\ion{Fe}{I}/H]$_{\mathrm{NLTE}}=-2.52\pm0.13$ and [\ion{Fe}{II}/H]$_{\mathrm{NLTE}}=-2.53\pm0.05$, whereas the LTE computation finds an average metallicity of ${\rm \overline{[Fe/H]}_{LTE}}\cong-2.57$. The quoted errors are the statistical line-to-line scatter; a solar iron abundance of $\log\epsilon(\mathrm{Fe})_{\sun}=7.45$ \citep[as recommended by][]{Asplund:2005} is assumed. Despite the fact that different model atmospheres were used for the determination of the stellar parameters and our abundance analysis, the LTE ionization equilibria of \ion{Fe}{I} and \ion{Fe}{II} in both stars indicate the consistency of the pressure stratification. The effective temperature was found through Balmer-line profile fits of both the $H\alpha$ and $H\beta$ lines to manually rectified spectra. The treatment of Stark broadening follows \citet{Vidaletal:1973}; self-broadening by hydrogen collisions is described using the recipe of \citet{AliGriem:1965}. A short summary of this method can be found in \citet{Korn:2004}; details of the calculations are given in \citet{Gehrenetal:2001a,Gehrenetal:2001b}, \citet{KornGehren:2002} and \citet{Kornetal:2003}. The interdependencies of the stellar parameters were resolved by corrections and iteration until convergence was achieved. The stellar parameters determined with this method are $T_{\mathrm{eff}}=5060\pm100$\,K, with a surface gravity of $\log g=2.3\pm0.2$ for \object{HE~1219$-$0312}; for \object{CS~29491$-$069}, the result is $T_{\mathrm{eff}}=5300\pm100$\,K and $\log g=2.8\pm0.2$. These are the stellar parameters that we adopted for our abundance analyses. Figure~\ref{fig:Isochrones} compares these parameters with a
$\mathrm{[Fe/H]}=-2.5$, 12\,Gyr isochrone of \citet{Yi:2001,Kim:2002}.

P.S.B. independently determined the effective temperatures of both stars by applying the Stark broadening theory described in \citet{StehleHutcheon:1999} and the self-broadening theory of \citet{BPO:2000}, using \texttt{MARCS} model atmospheres, and adopting HERES surface gravities and metallicities. Synthetic profiles fits were performed for both H$\alpha$ and H$\beta$, using an automated procedure \citep[see][]{Barklemetal:2002}. This approach yielded $T_{\mathrm{eff}}=5060$\,K for \object{HE~1219$-$0312}. Corrected for the higher HERES surface gravity ($\log g=2.4$), this translates into $T_{\mathrm{eff}}=5080$\,K at $\log g=2.3$, and is thus consistent with A.J.K.'s parameter determination.  For \object{CS~29491$-$069}, the analysis yielded $T_{\mathrm{eff}}=5160$\,K, assuming $\log g=2.8$ (which is the surface gravity derived by A.J.K.), and $T_{\mathrm{eff}}=5240$\,K, assuming $\log g=2.5$ (i.e., the surface gravity as published in Paper II).

The conclusion from a comparison of these results is that systematic discrepancies can reach up to $200$\,K in $T_{\mathrm{eff}}$ and $0.3$\,dex in $\log g$; the abundances are therefore not totally independent of the applied methods. Our adopted abundance uncertainties should give a reasonably realistic approximation though; the chemical similarity of most r-process elements that were studied in this work reduces the sensitivity of their abundance ratios to the stellar parameters (see Sect. \ref{sec:ErrorBudget}). Moreover, both stars reach perfect ionization equilibrium for \ion{Ti}{I}/\ion{Ti}{II} and a reasonable agreement for \ion{Fe}{I}/\ion{Fe}{II}.

\subsection{Model atmospheres}\label{sec:marcs}

Model atmospheres for both stars were computed with the latest version of the \texttt{MARCS} package \citep{Gustafssonetal:2008}. The \texttt{MARCS} models assume 1D plane-parallel stratification or spherical symmetry, depending on the
surface gravity, as well as hydrostatic equilibrium and radiative transfer in local thermodynamic equilibrium (LTE), also including continuum scattering.  Full equilibrium is computed for more than $600$ molecules. The various effects of convection are
approximated with the mixing length theory, a microturbulence parameter, and a term for turbulent pressure ($P_{\rm  turb}\propto \rho v_t^2$) in the hydrostatic equation. Energy conservation is fulfilled by assuming flux constancy for radiative and convective transport. Opacity sampling algorithms were included in the version of the \texttt{MARCS} code that we used, providing accurate depth-dependent opacities at approximately $10^5$ wavelength points between 1300\,{\AA} and 20\,$\mu$. Several model atmospheres were computed with varying chemical compositions, following the iterative process of abundance determinations, to account for the feedback of abundant elements such as C, N, O, Ca, or Fe on the electron pressure and the molecular equilibrium.

The impact of r-process enhancement on the atmospheric structure was also examined and found to be negligible. Overabundances of some $\alpha$-elements (e.g. O, Mg, or Ca), a common feature in metal-poor stars, were found in \object{CS~29491$-$069} and \object{HE~1219$-$0312}, and taken into account in the computation of the model atmospheres which were used in the final iteration of the abundance analysis.

\subsection{Line data}

Most of the line data for elements lighter than yttrium, which were adopted for the analysis, is identical to that used by \citet{Cayreletal:2004}, which in turn had mostly been taken from the Vienna Atomic Line Database (VALD; \citealt{VALD1,VALD2,Ryabchikovaetal:1999}). Line data for yttrium and heavier atoms are mostly identical with those adopted by \citet{Snedenetal:1996,Snedenetal:2000}, but were updated for lanthanum and europium \citep{Lawleretal:2001a,Lawleretal:2001b}, as well as for uranium and thorium \citep{Nilssonetal:2002a,Nilssonetal:2002b}. Recent transition probabilities and line positions for \element{Nd}, \element{Sm}, \element{Er} and \element{Hf} lines were also included. The data was taken from \citet{Denhartog:2003} and \cite{Lawleretal:2006,Lawleretal:2007,Lawleretal:2008}.

Hyperfine structure (HFS) was calculated for several important transitions: we use hyperfine constants from \citet{Davisetal:1971} for the ground state and \citet{Handrichetal:1969} for the $3d^54s4p\,z^{6}P$ excited state of three transitions of \ion{Mn}{I} around $4030$\,{\AA}. Further constants were taken from \citet{Holtetal:1999} for \ion{Mn}{II} \citep[$gf$-values from][]{Martinsonetal:1977}, \citet{Rutten:1978} compiled data for \element{Ba}, \citet{Lawleretal:2001a} give measurements for \element{La}, \citet{Ivarssonetal:2001} for \element{Pr}, and \citet{Lawleretal:2001b,Lawleretal:2004} for \element{Eu} and \element{Ho}.

Molecular line data for four different species (\element{CH}, \element{NH}, \element{OH} and \element{CN}) were assembled for the abundance measurements of the CNO elements. A \element{CH} line list was provided by B. Plez (2006, priv. comm.): The $gf$-values and line positions were taken from LIFBASE \citep{Luque/Crosley:1999}, and the excitation energies from \citet{Jorgensenetal:1996}. Isotopic shifts between \element[][12]{CH} and \element[][13]{CH} were computed by B. Plez.

The \element{NH} molecular line data was taken from \citet{KuruczCD:1993}. \citet{Aokietal:2006} found the line data to fit the Sun if a correction of the $gf$-values by $-0.4$\,dex (equivalent to a correction of the \element{N} abundance by $+0.4$\,dex) is applied; we have followed their recommendation. The \element{CN} line list was provided by B. Plez (2001, priv. comm.).

\subsection{Equivalent width analysis}\label{sec:EqwAnalysis}

Abundance analysis based on equivalent width measurements is a common method for determining the chemical composition of metal-poor stars, as blends are much less severe compared to more metal-rich stars. A large number of practically unblended spectral lines was identified and measured in the spectra of both stars. However, this method becomes less practical, or fails in the presence of HFS, stronger blends, or for analyzing molecular features. These cases were treated with spectrum
synthesis. Equivalent widths, $W_{\lambda}$, were determined by a simultaneous $\chi^2$ fit of a Gaussian and a straight line, representing the line profile and the continuum, respectively. The choice of a Gaussian profile limits the method to weak lines dominated by Doppler cores. Spectrum synthesis was performed for saturated lines, such as the \ion{Si}{I} transition at $3905.5$\,{\AA}.

Given the comparatively high noise level of the spectra in the blue and UV regions, where most of the lines are found, we fitted the continuum with an automated procedure that provides for a more objective continuum placement. Regions where the true continuum is hidden by numerous weak absorption lines can often hardly be distinguished from pure noise. This leads to systematic errors which consequently increase scatter among the line abundances.

The continuum was found through iterative parabolic fits to the observed flux in a spectral window of typically 10\,{\AA}, carefully avoiding the damping wings of highly saturated lines or very crowded regions. All data points belonging to absorption lines are subsequently suppressed in each iteration with a $\kappa$-$\sigma$-clipping method, where $\sigma$ denotes the amplitude of the photon noise and $\kappa$ is a threshold value. Convergence is typically achieved after a few iterations when the number of data points assumed as true continuum remains constant. The threshold $\kappa$ parameter was determined empirically and then set to a fixed value. Line selection was performed manually to avoid misidentifications and to visually verify possible blends.

The equivalent width measurements were then translated into abundances by computation of synthetic widths. We used the \texttt{eqwi} program (versions 07.03/07.04) and the respective \texttt{MARCS} model atmosphere to calculate line formation in LTE (including continuum scattering as well) and spherical symmetry. A set of 77 \ion{Fe}{I} (\object{CS~29491$-$069}) as well as 45 \ion{Fe}{I} and 18 \ion{Ti}{II} lines (\object{HE~1219$-$0312}) were used to determine the microturbulence parameters $\xi_\mathrm{CS29491}=1.6\pm0.2$\,km/s and $\xi_\mathrm{HE1219}=1.6\pm0.1$\,km/s, by requiring that the derived abundances be independent of the line strength.

\begin{table*}[htbp]
 \caption{\label{tab:AbundanceSummary} LTE abundances of \object{CS~29491$-$069} and \object{HE~1219$-$0312} and error estimates$^{1}$.}
 \centering
  \begin{tabular}{r l l | r r r r r r | r r r r r r }\hline\hline
   & & & \multicolumn{6}{ c |}{\object{CS~29491$-$069}} & \multicolumn{6}{c}{\object{HE~1219$-$0312}}\\
   Z & Species & $\log\epsilon_{\sun}$ & N$^{2}$ & $\log\epsilon$ & $\sigma_{\log\epsilon}$ & $\sigma_\mathrm{tot}$ & synt & $[\textrm{X/Fe}]^{3}$
   & N$^{2}$ & $\log\epsilon$ & $\sigma_{\log\epsilon}$ & $\sigma_\mathrm{tot}$ & synt & $[\textrm{X/Fe}]^{3}$\\[0.1cm]\hline 
    6 & CH    & 8.39 &  1 & $ 6.11$ &    --  & $0.25$ & $+$ &  $ 0.23$ &  2 & $ 5.46$ & $0.02$ & $0.23$ & $+$ & $ 0.03$ \\
    7 & CN/NH & 7.78 &  1 & $ 5.50$ &    --  & $0.43$ & $+$ &  $ 0.23$ &  1 & $ 5.10$ &    --  & $0.35$ & $+$ & $ 0.28$ \\
    8 & OH    & 8.66 & -- &     --  &    --  &    --  & --  &      --  &  4 & $ 6.20$ & $0.16$ & $0.32$ & $+$ & $ 0.50$ \\
   11 & Na I  & 6.17 & -- &     --  &    --  &    --  & --  &      --  &  2 & $ 2.99$ & $0.10$ & $0.15$ & --  & $-0.22$ \\
   12 & Mg I  & 7.53 &  4 & $ 5.33$ & $0.28$ & $0.29$ & --  &  $ 0.31$ &  6 & $ 4.98$ & $0.19$ & $0.22$ & --  & $ 0.41$ \\
   13 & Al I  & 6.37 &  1 & $ 3.87$ &    --  & $0.19$ & --  &  $ 0.01$ &  1 & $ 3.16$ &    --  & $0.19$ & --  & $-0.25$ \\
   14 & Si I  & 7.51 &  1 & $ 5.20$ &    --  & $0.27$ & $+$ &  $ 0.20$ &  1 & $ 4.84$ &    --  & $0.31$ & $+$ & $ 0.29$ \\
   20 & Ca I  & 6.31 & 10 & $ 4.02$ & $0.15$ & $0.18$ & --  &  $ 0.22$ &  6 & $ 3.46$ & $0.14$ & $0.17$ & --  & $ 0.11$ \\
   21 & Sc II & 3.05 &  7 & $ 0.78$ & $0.13$ & $0.16$ & --  &  $ 0.24$ &  4 & $ 0.09$ & $0.06$ & $0.11$ & --  & $ 0.00$ \\
   22 & Ti I  & 4.90 & 11 & $ 2.73$ & $0.11$ & $0.16$ & --  &  $ 0.34$ &  7 & $ 2.08$ & $0.02$ & $0.13$ & --  & $ 0.14$ \\
   22 & Ti II & 4.90 &  9 & $ 2.73$ & $0.12$ & $0.15$ & --  &  $ 0.34$ & 18 & $ 2.08$ & $0.09$ & $0.13$ & --  & $ 0.14$ \\
   23 & V I   & 4.00 &  5 & $ 1.58$ & $0.19$ & $0.21$ & --  &  $ 0.09$ & -- &    --   &    --  &    --  & --  & --      \\
   23 & V II  & 4.00 &  7 & $ 1.57$ & $0.14$ & $0.30$ & --  &  $ 0.08$ &  3 & $ 0.79$ & $0.09$ & $0.14$ & --  & $-0.25$ \\
   24 & Cr I  & 5.64 &  5 & $ 2.86$ & $0.07$ & $0.15$ & --  &  $-0.27$ &  6 & $ 2.37$ & $0.12$ & $0.18$ & --  & $-0.31$ \\
   25 & Mn I  & 5.39 &  3 & $ 2.11$ & $0.03$ & $0.17$ & HFS &  $-0.77$ &  3 & $ 1.60$ & $0.03$ & $0.17$ & HFS & $-0.83$ \\
   25 & Mn II & 5.39 &  2 & $ 2.21$ & $0.09$ & $0.16$ & HFS &  $-0.67$ &  2 & $ 2.01$ & $0.06$ & $0.16$ & HFS & $-0.42$ \\
   26 & Fe I  & 7.45 & 77 & $ 4.90$ & $0.12$ & $0.18$ & --  &  $-0.04$ & 45 & $ 4.53$ & $0.09$ & $0.16$ & --  & $ 0.04$ \\
   26 & Fe II & 7.45 & 16 & $ 4.94$ & $0.14$ & $0.16$ & --  &  $ 0.00$ &  7 & $ 4.49$ & $0.11$ & $0.14$ & --  & $ 0.00$ \\
   27 & Co I  & 4.92 &  6 & $ 2.68$ & $0.12$ & $0.19$ & --  &  $ 0.27$ &  4 & $ 2.13$ & $0.07$ & $0.16$ & --  & $ 0.17$ \\
   28 & Ni I  & 6.23 &  8 & $ 3.66$ & $0.19$ & $0.25$ & --  &  $-0.06$ &  4 & $ 3.27$ & $0.06$ & $0.17$ & --  & $ 0.00$ \\
   30 & Zn I  & 4.60 &  2 & $ 2.23$ & $0.05$ & $0.08$ & --  &  $ 0.14$ &  1 & $ 1.70$ &    --  & $0.10$ & --  & $ 0.06$ \\
   38 & Sr II & 2.92 &  2 & $ 0.56$ & $0.07$ & $0.18$ & --  &  $ 0.15$ &  2 & $ 0.31$ & $0.04$ & $0.18$ & --  & $ 0.35$ \\
   39 & Y II  & 2.21 & 14 & $-0.24$ & $0.11$ & $0.15$ & --  &  $ 0.06$ & 13 & $-0.48$ & $0.11$ & $0.15$ & --  & $ 0.27$ \\
   40 & Zr II & 2.59 &  8 & $ 0.52$ & $0.13$ & $0.17$ & --  &  $ 0.44$ &  6 & $ 0.28$ & $0.09$ & $0.14$ & --  & $ 0.65$ \\
   46 & Pd I  & 1.69 &  -- &  --  &    --  &  --  & --  &   --  &  1 & $-0.25$ &    --  & $0.23$ & --  & $ 1.02$ \\
   56 & Ba II & 2.17 &  4 & $-0.10$ & $0.07$ & $0.12$ & HFS &  $ 0.24$ &  2 & $-0.14$ & $0.03$ & $0.13$ & HFS & $ 0.65$ \\
   57 & La II & 1.13 &  3 & $-0.86$ & $0.04$ & $0.12$ & HFS &  $ 0.52$ &  3 & $-0.86$ & $0.05$ & $0.12$ & HFS & $ 0.97$ \\
   58 & Ce II & 1.58 &  6 & $-0.28$ & $0.12$ & $0.16$ & --  &  $ 0.65$ & 13 & $-0.52$ & $0.18$ & $0.21$ & --  & $ 0.86$ \\
   59 & Pr II & 0.71 &  1 & $-0.95$ &    --  & $0.19$ & HFS &  $ 0.85$ &  3 & $-1.17$ & $0.05$ & $0.14$ & HFS & $ 1.08$ \\
   60 & Nd II & 1.45 &  8 & $-0.27$ & $0.09$ & $0.14$ & --  &  $ 0.79$ &  7 & $-0.41$ & $0.04$ & $0.12$ & --  & $ 1.10$ \\
   62 & Sm II & 1.01 &  4 & $-0.47$ & $0.13$ & $0.17$ & --  &  $ 1.03$ &  6 & $-0.59$ & $0.11$ & $0.15$ & --  & $ 1.36$ \\
   63 & Eu II & 0.52 &  4 & $-1.03$ & $0.02$ & $0.10$ & HFS &  $ 0.96$ &  3 & $-1.06$ & $0.03$ & $0.10$ & HFS & $ 1.38$ \\
   64 & Gd II & 1.12 &  8 & $-0.25$ & $0.05$ & $0.12$ & --  &  $ 1.14$ & 10 & $-0.41$ & $0.10$ & $0.15$ & --  & $ 1.43$ \\
   66 & Dy II & 1.14 &  8 & $-0.22$ & $0.05$ & $0.13$ & --  &  $ 1.15$ & 12 & $-0.34$ & $0.07$ & $0.13$ & --  & $ 1.48$ \\
   67 & Ho II & 0.51 &  1 & $-1.02$ &    --  & $0.21$ & HFS &  $ 0.98$ &  2 & $-1.13$ & $0.06$ & $0.17$ & HFS & $ 1.32$ \\
   68 & Er II & 0.93 &  5 & $-0.38$ & $0.07$ & $0.13$ & --  &  $ 1.20$ &  2 & $-0.56$ & $0.07$ & $0.14$ & --  & $ 1.47$ \\
   69 & Tm II & 0.00 & -- &     --  &    --  &    --  & --  &      --  &  1 & $-1.51$ &    --  & $0.16$ & $+$ & $ 1.45$ \\
   72 & Hf II & 0.88 &  1 &     --  &    --  &    --  & --  &      --  &  1 & $-0.97$ &    --  & $0.22$ & $+$ & $ 1.11$ \\
   76 & Os I  & 1.45 &  1 & $<0.03$ &    --  &    --  & $+$ &  $<1.09$ & -- &     --  &    --  &    --  & --  &    --   \\
   90 & Th II &  0.06 &  1 & $-1.43$ &    --  & $0.22$ & $+$ &    1.02  &  1 & $-1.29$ &    --  & $0.14$ & $+$ &   1.61  \\[0.1cm]
   \hline
   \multicolumn{15}{l}{1: see Sect.~\ref{sec:ErrorBudget}; 2: number of detected lines or molecular bands; 3: \ion{Fe}{II} was chosen as the reference iron abundance}
   \end{tabular}
\end{table*}

\subsection{Spectrum synthesis}\label{sec:SpectrumSynthesis}

The spectrum synthesis method was applied whenever strong lines or features with multiple transitions, such as blends, molecular features, or hyperfine structure (HFS), had to be analyzed. Computations of line formation were conducted using the \texttt{bsyn} program and the respective \texttt{MARCS} model atmospheres. The synthetic spectra were convolved with a Gaussian function, representing the instrumental profile $\Gamma$. The FWHM depends on the slit width, and was determined using the atlas of weak ThAr lines in the UVES calibration data (see the resolving power, $R=\lambda/\Delta\lambda$, in Table~\ref{tab:Observation}). While the result matched the spectra of \object{CS~29491$-$069} ($R_{\mathrm{med}}=57,600$; $\Gamma_{\mathrm{med}}=5.2$\,km/s), additional broadening in \object{HE~1219$-$0312} increased the profile width to $\Gamma=6.1$\,km/s. This may be due to macroscopic gas movements such as convection. In most cases, the instrumental profile was set to a fixed FWHM to provide for more consistent fits of weaker lines, since the noise level of the spectra often prevented a simultaneous profile fit. In some cases, e.g. lines which are located near the edge of an echelle order, mismatches of synthetic and measured fluxes were resolved by slight adjustments of the profile width.

A window of $\sim50$\,{\AA} around the region of interest was rectified using the same continuum placement method as described above. We then applied $\chi^2$ fits of the synthetic profile, where $\chi^2$ varies with the abundance, $\log\epsilon$, to improve fits which were affected by high noise levels. The $1\sigma$ errors representing the photon noise in the detector were used for computing the fit weights.

\begin{figure*}[t!]
 \centering
  \includegraphics[width=12cm]{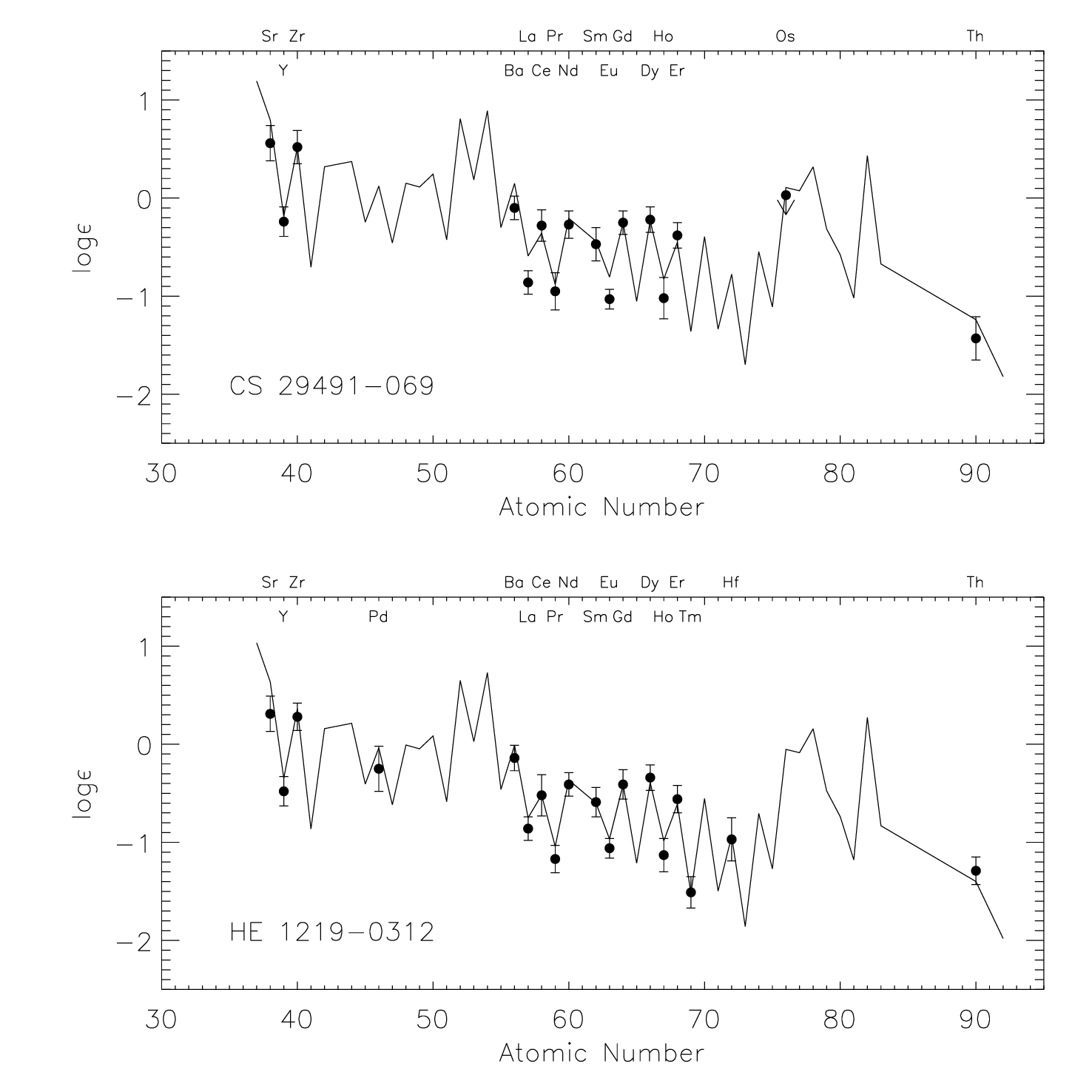}
  \caption{\label{Fig:abundance_patterns} Heavy-element abundance 
    patterns of \object{CS~29491$-$069} (upper panel) and \object{HE~1219$-$0312} 
    (lower panel) compared to solar residuals which were not formed by the s-process \citep[][see Sect. \ref{sec:heavyelm}]{Arlandinietal:1999}, scaled to
    match the observed Gd abundance.}
\end{figure*}

\begin{figure*}[t!]
 \centering
  \includegraphics[width=12cm]{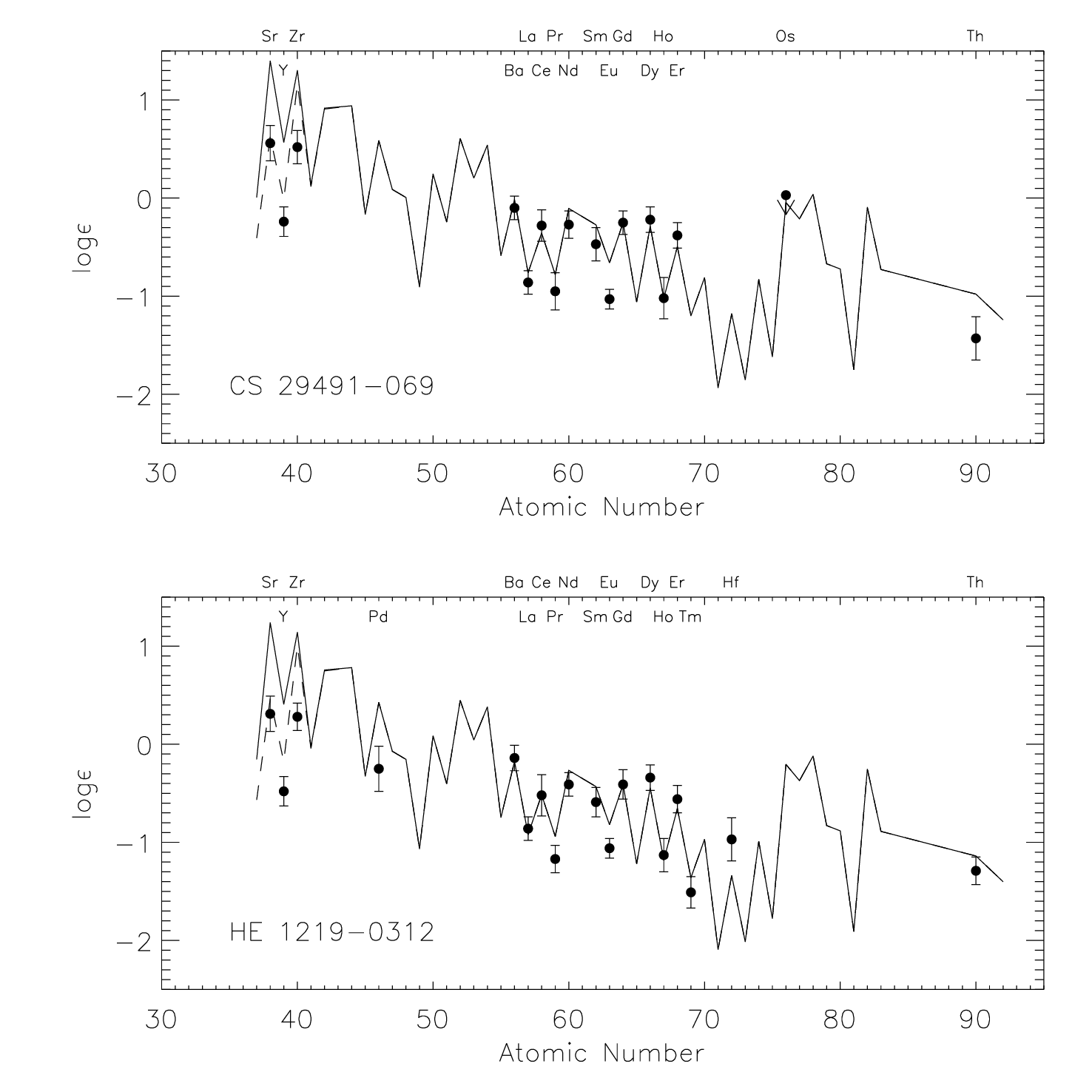}
  \caption{\label{Fig:abundance_patternsHEW} Heavy-element abundance 
    patterns of \object{CS~29491$-$069} (upper panel) and \object{HE~1219$-$0312} 
    (lower panel) compared to solar-like HEW scenario abundance yields for an entropy range of $10\le S\le280$ (straight line) and $60\le S\le280$ (dashed line), scaled to
    match the observed Gd abundance. See Sect. \ref{sec:NcapHEW} for details.}
\end{figure*}

\section{Results}\label{sec:Results}

Table~\ref{tab:AbundanceSummary} gives an overview of the abundances of \object{CS~29491$-$069} and \object{HE~1219$-$0312}. Below we comment on individual groups of elements and how our findings compare to the other HERES stars presented in Paper II, as well as a sample of similar stars found in \citet{Cayreletal:2004}.

\subsection{CNO}\label{sec:CNO}

The carbon abundance is very important for accurate measurements of both thorium and uranium, as their most important features, \ion{Th}{II} $4019$\,{\AA} and \ion{U}{II} $3860$\,{\AA}, are blended with \element[][13]{CH} and \element{CN} lines, respectively. In both stars, the C abundance could easily be determined through spectrum synthesis of the \element{CH} G-band at $4300$\,{\AA}. While \object{HE~1219$-$0312} is not carbon-enhanced ($\mathrm{[C/Fe]}\sim0.0$), \object{CS~29491$-$069} shows slight carbon enhancement of $+0.23$\,dex, although consistent with zero overabundance within the measurement uncertainties. The relative \element{C} abundances of both stars are similar to those of most other HERES stars at these metallicities and in the sample of \citet{Cayreletal:2004}.

The CH line lists were also applied in an attempt to measure $\element[][12]{C}/\element[][13]{C}$ isotopic ratios of the stars. However, the limited $S/N$ of the spectrum of \object{HE~1219$-$0312} around the G-band only permitted a determination of a lower limit of $\element[][12]{C}/\element[][13]{C}\gtrsim 6$. The spectrum of \object{CS~29491$-$069} has slightly better $S/N$; the lower limit for the isotopic ratio is slightly higher, $\element[][12]{C}/\element[][13]{C}\gtrsim 10$. No clear traces of \element[][13]{CH} could be detected between the \ion{Nd}{II} and \ion{Th}{II} features (see Fig. \ref{Fig:ThoriumSynthesis} and Table \ref{tab:ThII4019linelist} for the line positions), by virtue of comparatively low carbon abundances found in both stars. We therefore adopted solar $\element[][12]{C}/\element[][13]{C}\gtrsim 90$ for the measurement of \ion{Th}{II} $4019$\,{\AA} to obtain good fits to the observed spectra.

Nitrogen was measured using the \element{NH} band at $3360$\,{\AA} for \object{HE~1219$-$0312}, while the $(0,0)$ band head of the \element{CN} B--X system at 3883\,{\AA} was used for \object{CS~29491$-$069}. The derived relative abundances in both stars are consistent with the Sun within the measurement uncertainties. The latter are rather high, owing to the weak line strength and low $S/N$. A strong \ion{Fe}{I} blend in the \element{CN} feature and the additional uncertainty of the carbon abundance further increases the error bars in \object{CS~29491$-$069}.

Oxygen could only be detected in \object{HE~1219$-$0312} using four features of the \element{OH} A--X band in the UV close to $3200$\,{\AA}. However, the low $S/N$ in this spectral range again limits the accuracy of this abundance measurement. The derived oxygen abundance of $\log\epsilon=6.20$ corresponds to an enhancement of $\mbox{[O/Fe]}=+0.50$. Such a value is expected given the observed overabundance of $\mbox{[O/Fe]}\sim +0.4$ in stars of similar metallicity \citep[e.g. ][]{Cayreletal:2004,Spiteetal:2005}, although the $1\sigma$ uncertainty is rather large. However, \citet{Colletetal:2007} predict corrections of up to $-0.96$\,dex for abundances derived with \element{OH} lines (at $0.5$\,eV excitation level and for a model with $T_\mathrm{eff}=5128$\,K, $\log g=2.2$, [Fe/H]=$-3.0$) using 3D time-dependent hydrodynamic model atmospheres. This is mainly attributed to the lower temperatures in the surface layers of 3D models, enhancing molecule formation. In this picture, \object{HE~1219$-$0312} would appear more oxygen-poor than stars of similar metallicity analyzed in \citet{Cayreletal:2004} using the forbidden \ion{O}{I} line at $6300$\,{\AA}. Only one \element{OH} feature employed for the analysis has an excitation level of $0.5$\,eV, though, the others range between $0.7$\,eV and $1.3$\,eV, which reduces the abundance corrections. \citet{Colletetal:2007} also caution that the treatment of scattering in their models needs refinement; Rayleigh scattering in the continuum is particularly important in the UV where the \element{OH} lines that were employed in the analysis are found. Considering the large uncertainty of the oxygen abundance, it remains unclear whether \object{HE~1219$-$0312} is oxygen-poor or not.

\subsection{Sodium to titanium}

Equivalent-width analysis and spectrum synthesis were conducted for the measurement of several elements in this group. \object{HE~1219$-$0312} exhibits clear enhancement of the $\alpha$-element \element{Mg}, compared to the solar Mg/Fe ratio, possibly also \element{Si}, as well as sub-solar abundances of \element{Na} and \element{Al} relative to Fe. This may be attributed to enrichment of the progenitor gas cloud by a type-II supernova. The magnesium and silicon abundances of \object{HE~1219$-$0312} are in the range covered by the sample of \citet{Cayreletal:2004}; the observed magnesium overabundance is similar to other HERES stars.

Applying an NLTE correction of $-0.5$\,dex to the sodium abundance, following \citet{Cayreletal:2004}, who used the same \ion{Na}{I} resonance lines for their analysis, leaves \object{HE~1219$-$0312} comparatively under-abundant. Contrary to that, adopting their NLTE correction of $+0.65$\,dex for Al leads to a relative overabundance. The same holds for the HERES sample; most stars exhibit significantly less aluminium.

The elements \element{Ca} and \element{Sc} follow the scaled solar abundances within their measurement uncertainties; \element{Ti} seems slightly over-abundant. All three elements lie in the lower abundance range of the HERES and \citet{Cayreletal:2004} samples.

The respective pattern in \object{CS~29491$-$069} at most exhibits a slight overabundance of \element{Mg} and no significant enhancement of \element{Al} and \element{Si}, these elements reflect the scaled solar abundance pattern within the error bars. Magnesium does not exhibit any unusual behavior compared to the HERES and \citet{Cayreletal:2004} samples; silicon is at the lower end of the range of \citet{Cayreletal:2004}. The \element{Al} overabundance compared to both samples is even more pronounced than in the case of \object{HE~1219$-$0312}. The number of stars at [Fe/H]$\sim-2.5$ in \citet{Cayreletal:2004} is small, however, and NLTE effects may not be properly accounted for in \object{CS~29491$-$069}, which has a higher effective temperature and surface gravity than most stars in this sample.

Silicon is under-abundant compared to the \citet{Cayreletal:2004} sample, but again there are only few other stars with similar metallicities. \element{Ca} and \element{Sc} seem enhanced in \object{CS~29491$-$069}; both elements lie at the extreme ends of the HERES and \citet{Cayreletal:2004} data sets.

Titanium exhibits clear overabundance ($\mathrm{[Ti/Fe]}=+0.34$), \object{CS~29491$-$069} seems more enhanced than the other stars in both samples.

\subsection{Iron-group elements}

A number of vanadium and chromium lines were detected in the spectra of both stars. \object{HE~1219$-$0312} seems more vanadium-poor than the other HERES stars, while \object{CS~29491$-$069} is not unusual. The Cr under-abundances of $-0.31$ (\object{HE~1219$-$0312}) and $-0.27$ (\object{CS~29491$-$069}) agree with the comparison samples.

Features arising from \ion{Mn}{I} and \ion{Mn}{II} were detected in both stars. We include HFS broadening in the analysis, which consequently strongly reduces the line-to-line abundance scatter.

We observe an abundance difference of $0.4$\,dex between the results for \ion{Mn}{I} and the two \ion{Mn}{II} lines in \object{HE~1219$-$0312}. The same discrepancy was also reported by various authors \citep{Johnson:2002,Cayreletal:2004,Jonselletal:2006} for other metal-poor stars, who attributed it to shortcomings in the structure of the stellar model atmosphere, uncertainties in the $\log gf$ values, or NLTE effects (the three features around $4030$\,{\AA} are resonance lines). Assuming the correction of $+0.4$\,dex for \ion{Mn}{I} adopted by \citet{Cayreletal:2004} leads to very good agreement between the \ion{Mn}{I} and \ion{Mn}{II} abundances; the observed under-abundance of about $-0.4$\,dex also agrees with their findings and the HERES stars.

In contrast, \ion{Mn}{I} and \ion{Mn}{II} in \object{CS~29491$-$069} differ by only 0.1\,dex; it is under-abundant in manganese compared to both samples.

A large number of \ion{Fe}{I} and \ion{Fe}{II} lines were employed to determine the microturbulence $\xi$. Adopting a solar iron abundance of $\log\epsilon(\mathrm{Fe})_{\sun}=7.45$ \citep{Asplund:2005}, the relative iron
abundances of \object{CS~29491$-$069} and \object{HE~1219$-$0312} are $\mathrm{[Fe/H]}=-2.51\pm0.16$ and $\mathrm{[Fe/H]}=-2.96\pm0.14$, respectively, as determined from \ion{Fe}{II} lines \citep[\ion{Fe}{II} is believed to be a more reliable iron abundance indicator; see][]{Asplund:2005b}.

Detections of \element{Co}, \element{Ni} and \element{Zn} lines complete the abundance determinations around the iron peak. \object{HE~1219$-$0312} and \object{CS~29491$-$069} follow the HERES sample and \citet{Cayreletal:2004}. However, \object{HE~1219$-$0312} is very \element{Zn}-poor compared to the HERES stars (although there are only few stars available at this metallicity with measured Zn abundances) and slightly under-abundant compared to the \citet{Cayreletal:2004} sample. 

\subsection{Heavy elements}\label{sec:heavyelm}

\begin{figure}[tbp]
  \centering
  \resizebox{\hsize}{!}{\includegraphics{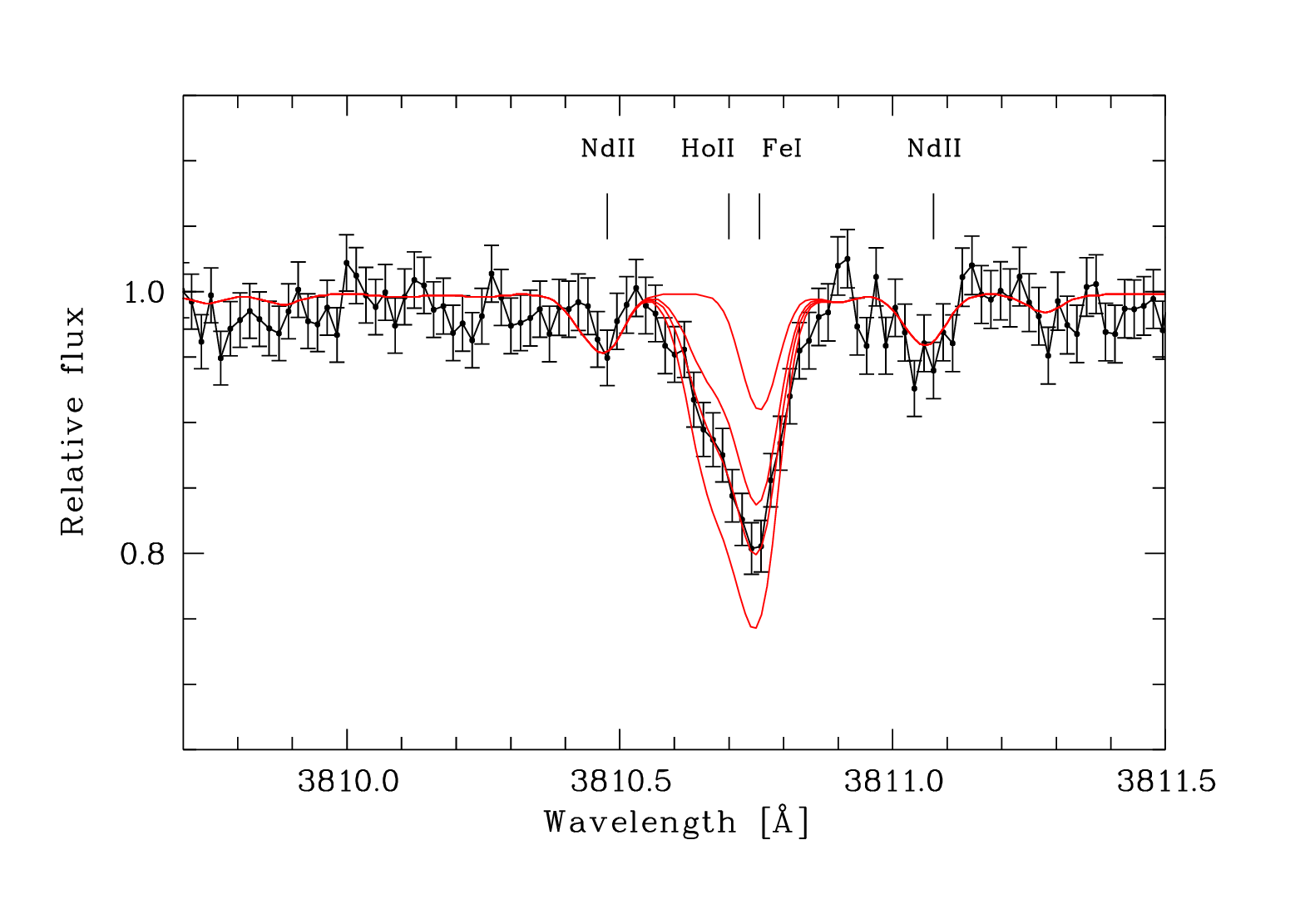}}
  \resizebox{\hsize}{!}{\includegraphics{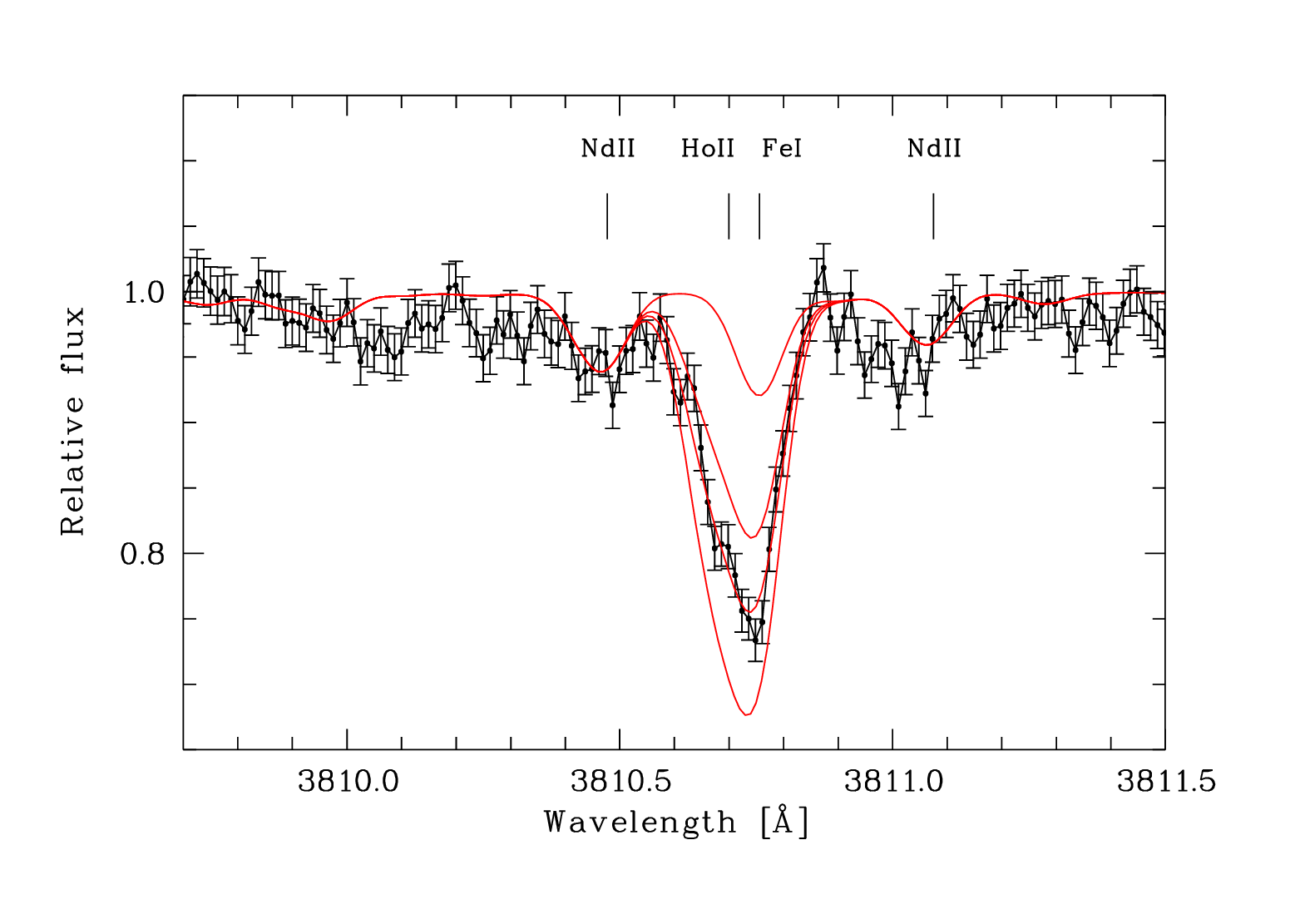}}
  \caption{\label{Fig:HolmiumSynthesis} Spectrum synthesis of the \ion{Ho}{II}
    3811\,{\AA} line in \object{CS~29491$-$069} (upper panel) and
    \object{HE~1219$-$0312}. Shown are synthesis calculations for the best-fit
    values (i.e., $\log\epsilon(\mathrm{Ho})=-1.02$ and $-1.17$,
    respectively), $\pm 0.2$\,dex, and no holmium.} 
\end{figure}

\begin{figure}[bp]
  \centering
  \resizebox{8cm}{!}{\includegraphics{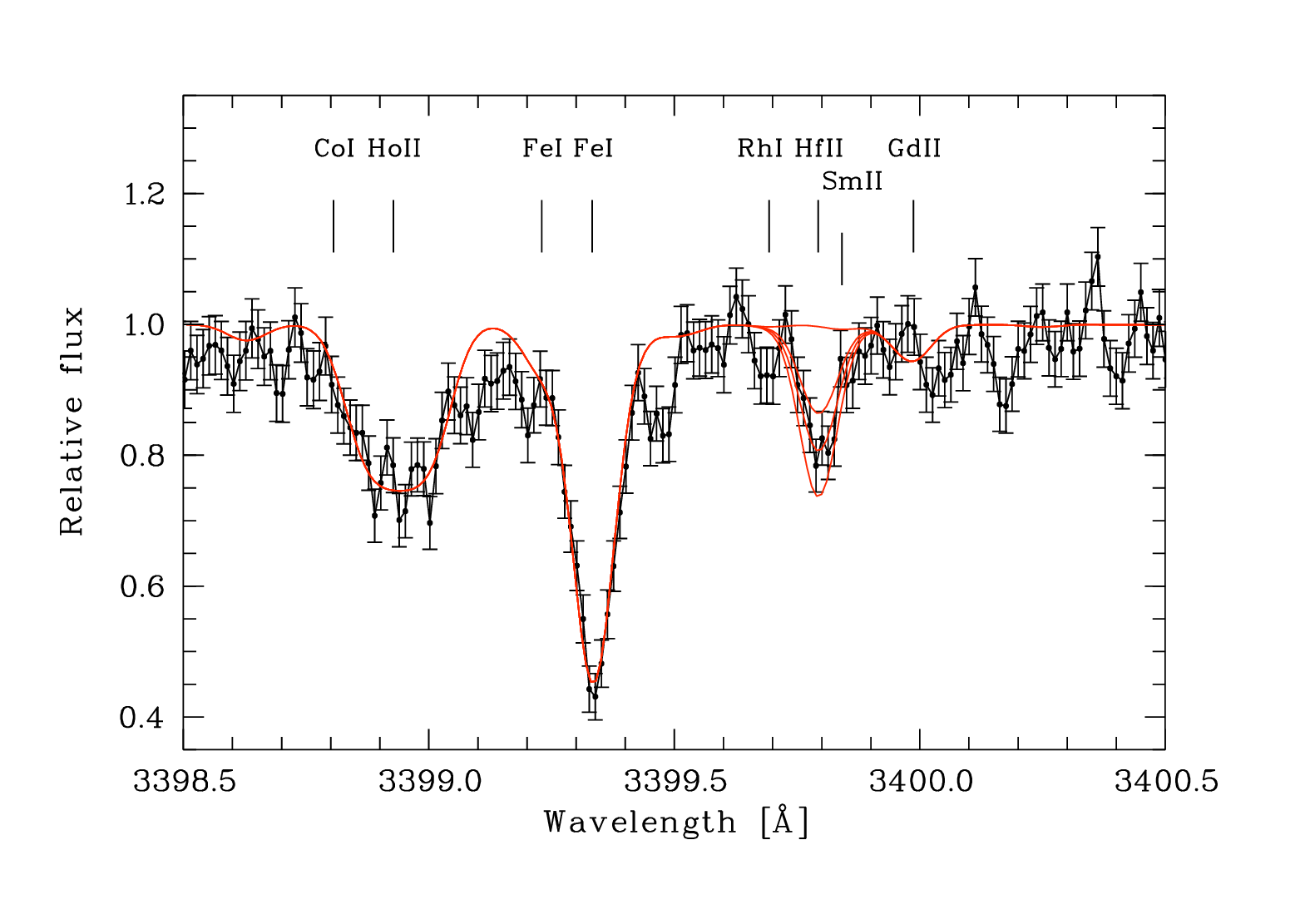}}
  \caption{\label{Fig:HafniumSynthesis} Spectrum synthesis of the \ion{Hf}{II}
    3400\,{\AA} line in \object{HE~1219$-$0312}. Shown are synthesis calculations for the best-fit
    values (i.e., $\log\epsilon(\mathrm{Hf})=-1.05$), $\pm 0.2$\,dex, and no hafnium.} 
\end{figure}

The outstanding feature of both stars is a high abundance of the heavy elements: $[\rm{r/Fe}]=+1.1$ (\object{CS~29491$-$069}) and $[\rm{r/Fe}]=+1.5$ (\object{HE~1219$-$0312}), where $r$ denotes an average of the abundances of europium (Eu), gadolinium (Gd), dysprosium (Dy) and holmium (Ho), pointing to their production by a rapid neutron-capture process. In Fig.~\ref{Fig:abundance_patterns}, we compare the heavy-element abundance patterns with scaled solar residual abundances not produced by the s-process. The decomposition of contributions is based on the total solar abundances of \citet{Asplund:2005} and the s-process fractions of \cite{Arlandinietal:1999}.

Among the light elements, Sr, Y, and Zr are easily accessible for measurements. Their origin has recently been under active discussion: studies of disk and halo stars by e.g. \citet{Aokietal:2005}, \citet{Mashonkinaetal:2007}, \citet{Francoisetal:2007} and \citet{Montesetal:2007} found that their formation cannot be explained by a simple split into contributions from the ``classical'' s-process and r-process. An anti-correlation between the [Sr/Ba], [Y/Ba] and [Zr/Ba] ratios against the barium abundance (in the range $-4.5\simeq[\rm{Ba/H}]\simeq-1.5$) was observed in \citet{Francoisetal:2007}. \citet{Montesetal:2007} found similar anti-correlations for [Sr/Eu], [Y/Eu] and [Zr/Eu] against [Eu/Fe] using abundance data from the literature. The authors of both publications confirm the existence of an additional lighter element primary process \citep[LEPP, ][]{Travaglioetal:2004}.

\object{CS~29491$-$069} and \object{HE~1219$-$0312} agree with the \citet{Montesetal:2007} Sr, Y and Zr abundance distributions. Both objects are also consistent with the solar residuals of Sr, Y and Zr when scaled to fit the heavier elements between the second and third r-process peaks. However, yttrium exhibits a significant under-abundance in both objects when compared to the r-process fractions of \citet{Burrisetal:2000}. The Sr and Y slopes in \citet{Montesetal:2007} exhibit a flattening towards the highest europium overabundances where \object{HE~1219$-$0312} is found, indicating that strong r-process enrichment may also contribute significantly to the production of these elements.

The relative abundance ratios [Sr/Y], [Sr/Zr] and [Y/Zr] seem largely independent of metallicity; this strongly supports a scenario in which these elements are formed by the same process, apart from the above mentioned possible r-process contribution: \citet{Francoisetal:2007} found $\mathrm{[Y/Sr]}=-0.2\pm0.2$ over metallicity with rather low scatter, consistent with $\mathrm{[Y/Sr]}=-0.08$ in \object{HE~1219$-$0312} and $\mathrm{[Y/Sr]}=-0.09$ in \object{CS~29491$-$069}. However, \citet{Francoisetal:2007} caution a possible anti-correlation of [Y/Sr] vs. [Sr/H] at the lowest metallicities. \citet{Farouqietal:2008b} give an average observed ratio of $\mathrm{Sr/Y}=6.17\pm1.06$, taken from various analyses of halo stars at different metallicities and r-process enhancements. This translates into $\mathrm{[Y/Sr]}=-0.08\pm0.07$, which is in very good agreement with our measurement. Their high-entropy-wind (HEW) models include a charged-particle process as a candidate for the LEPP (see Sect. \ref{sec:NcapHEW}).

Palladium was detected in \object{HE~1219$-$0312} only through the \ion{Pd}{I} $3404.579$\,{\AA} line, the strongest available feature of this species. It is similarly under-abundant, relative to the solar residuals, as in other highly r-process enhanced stars \citep{Montesetal:2007}. The $3404.579$\,{\AA} line was not covered by our spectra of \object{CS~29491$-$069}; a reliable Pd abundance could not be measured.

The elements beyond the second r-process peak, beginning with barium, are largely consistent with a ``classical'' pure r-process scenario (see also Sect. \ref{sec:NcapHEW}). Most of them are easily accessible for spectroscopic measurements.

The Ba abundances were determined using spectrum synthesis, including HFS and isotope splitting. The adopted isotope mix follows the r-process-only fractions published by \citet{McWilliam:1998}; that is, 40\,\% of \element[][135]{Ba}, 32\,\% of \element[][137]{Ba}, and 28\,\% of \element[][138]{Ba} (the isotopes \element[][134]{Ba} and \element[][136]{Ba} are not produced by the r-process). \object{CS~29491$-$069} seems slightly less enhanced with barium.

The Eu abundance was determined from three (four) practically unblended \ion{Eu}{II} lines at $3819.672$\,{\AA}, $3907.107$\,{\AA}, $4129.725$\,{\AA} and $4205.042$\,{\AA} (\object{CS~29491$-$069} only). All of these lines exhibit significant HFS broadening and were therefore analyzed with spectrum synthesis. A solar isotope mix was adopted as recommended by \citet{Sneden:2002}. The line profiles could be fitted with very good agreement, and the line-to-line scatter is small. The average Eu abundance lies beneath its scaled solar residual in \object{CS~29491$-$069}.

Measurements of the remaining lanthanides, La, Ce, Pr, Nd, Sm, Gd, Dy, Ho, Er and Tm (\object{HE~1219$-$0312} only) complete the analysis of rare-earth elements (see Fig. \ref{Fig:HolmiumSynthesis} for holmium). With the exceptions of lanthanum and europium in \object{CS~29491$-$069}, they closely follow the solar residuals within the error bars, a feature which has so far been commonly observed in r-process enhanced metal-poor stars. The abundances seem to follow a slight downward trend towards lighter elements with respect to the residuals (see Sect. \ref{sec:ages}), although our results are not decisive in that respect.

We report the measurement of hafnium in \object{HE~1219$-$0312}, a stable transition element which is suitable for nuclear age determination due to its proximity to the third-peak elements \citep{Kratzetal:2007}; see the discussion in Sect. \ref{sec:ages}. We measure one transition at $3399.79$\,{\AA}; the low $S/N$ in the UV leads to rather large abundance uncertainty. The line is weakly blended with \ion{Sm}{II} (see Fig. \ref{Fig:HafniumSynthesis}). Hafnium is not detected in \object{CS~29491$-$069}, most likely due to its weaker r-process enhancement. We also measure an upper limit for the third-peak element osmium in \object{CS~29491$-$069}; this element is not detected in \object{HE~1219$-$0312}.

The lead abundance has attracted a lot interest in the last years in connection with the validity of r-process model yields and nuclear age dating. Lead isotopes lie in various decay chains, in particular those of thorium and uranium, and provide for an important test of the observed actinide abundances and their theoretical predictions (although this picture may be more complicated, see Sect. \ref{sec:ages}). Unfortunately, the quality of our spectra was not sufficient for measuring Pb abundances, due to the very small equivalent widths of \ion{Pb}{I} features. The noise level in the spectra did not allow the determination of a useful upper limit.

\begin{figure}[tbp]
  \centering
  \resizebox{8cm}{!}{\includegraphics{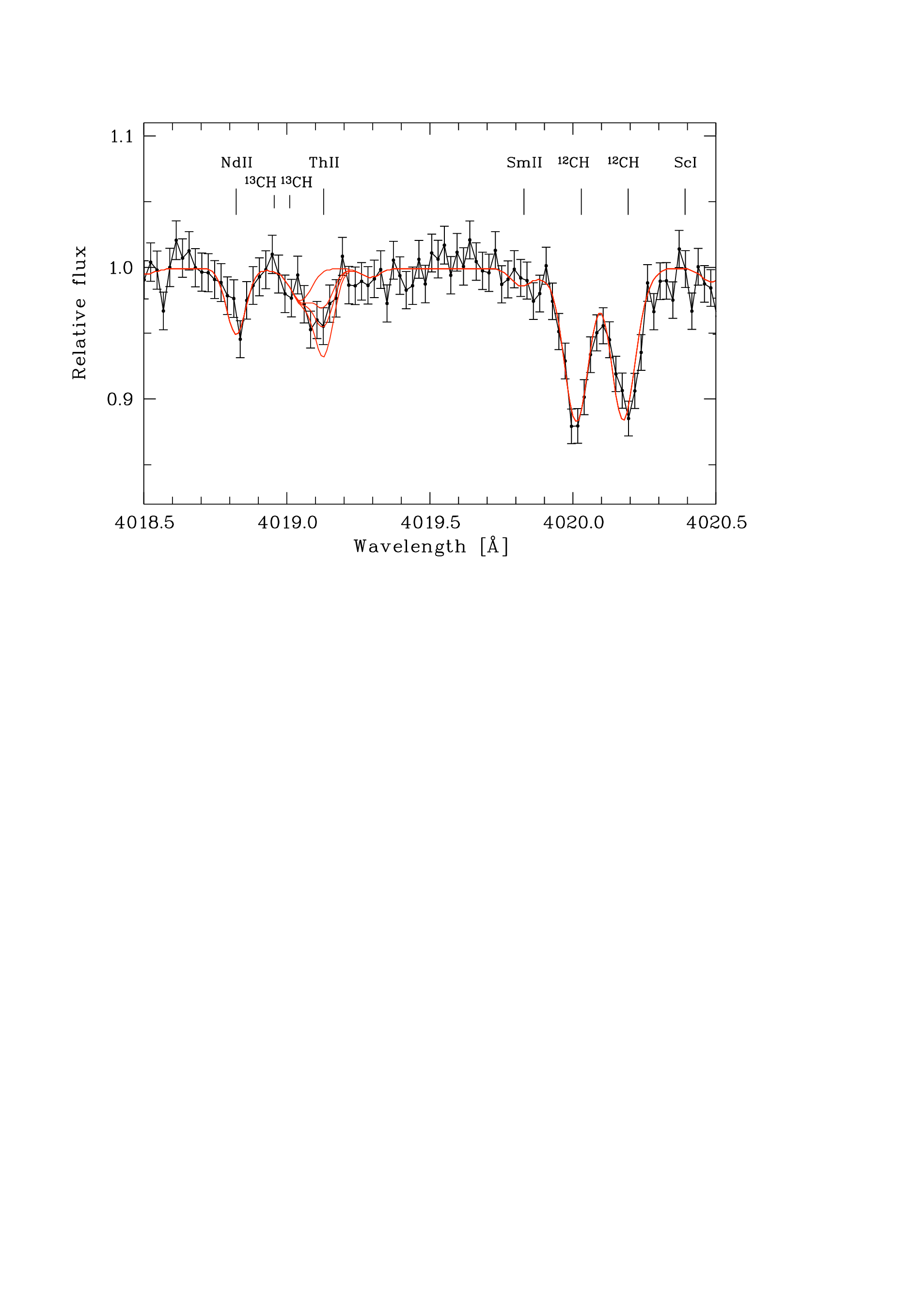}}
  \resizebox{8cm}{!}{\includegraphics{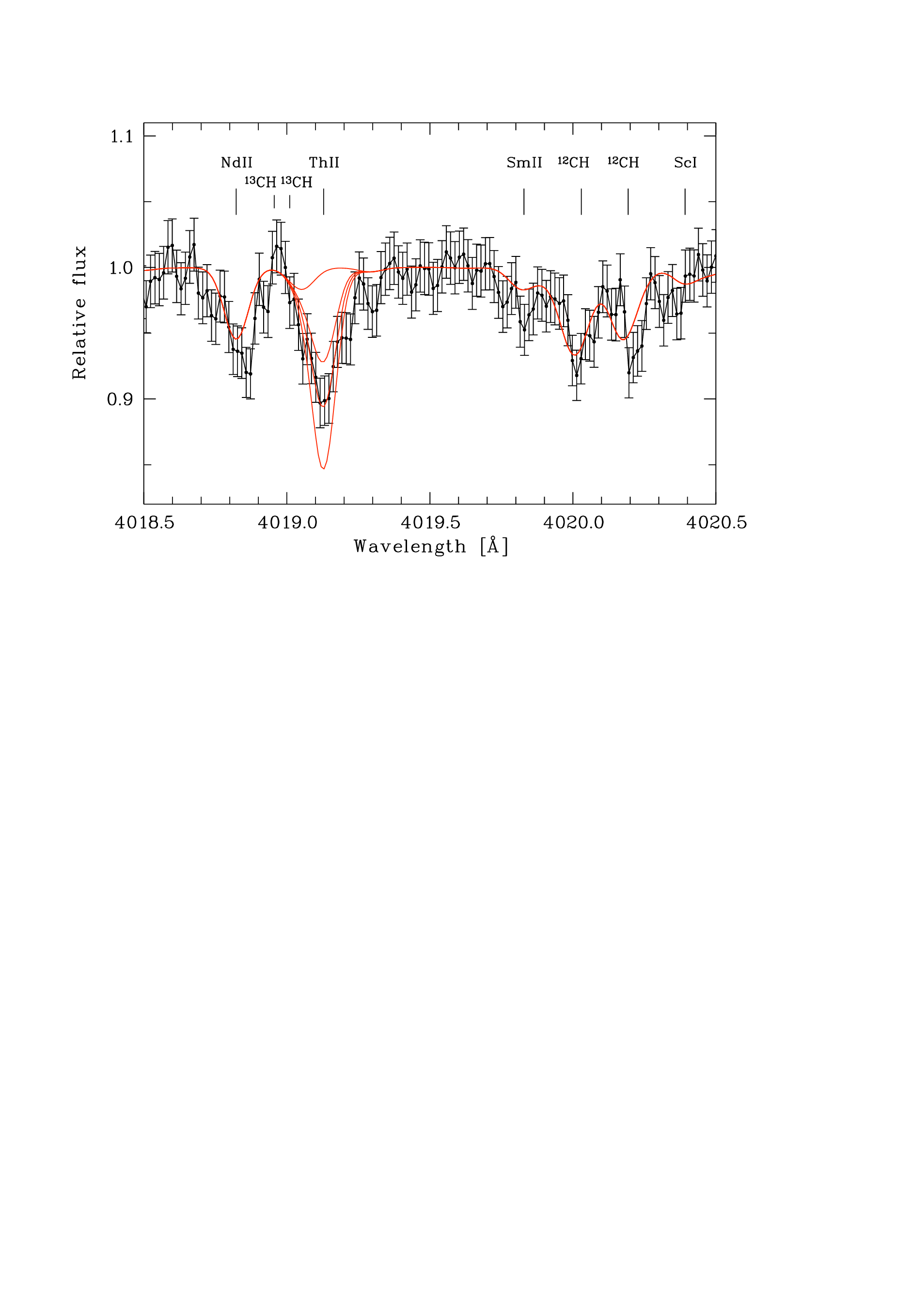}}
  \caption{\label{Fig:ThoriumSynthesis} Spectrum synthesis of the \ion{Th}{II}
    4019\,{\AA} line in \object{CS~29491$-$069} (upper panel) and
    \object{HE~1219$-$0312}. Shown are synthesis calculations for the best-fit
    values (i.e., $\log\epsilon(\mathrm{Th})=-1.43$ and $-1.29$,
    respectively), $\pm 0.2$\,dex, and no thorium.} 
\end{figure}

\begin{table}[bp]
\caption{Line list adopted for the spectrum synthesis of the \ion{Th}{II} 
  $4019.129$\,{\AA} line region.}
\begin{center}
\begin{tabular}{rccc}\hline\hline
Species			&       $\lambda_0$ [{\AA}]     &       $\chi_{e}$              &       $\log gf$\\
\hline
\ion{Fe}{I}			&       4018.506			&       4.209           &       $-1.597$\\
\element[][12]{CH}	&       4018.633			&       0.341           &       $-3.799$\\
\element[][13]{CH}	&       4018.704			&       0.733           &       $-7.272$\\
\ion{Nd}{II}		&       4018.823			&       0.064           &       $-0.850$\\
\element[][13]{CH}	&       4018.956			&       0.463           &       $-1.379$\\
\ion{Fe}{I}			&       4019.042			&       2.608           &       $-2.780$\\
\element[][13]{CH}	&       4019.010			&       0.463           &       $-1.354$\\
\ion{Ce}{II}		&       4019.057			&       1.014           &       $-0.213$\\
\ion{Ni}{I}			&       4019.058			&       1.935           &       $-3.174$\\
\ion{Th}{II}			&       4019.129			&       0.000           &       $-0.228$\\
\ion{Co}{I}			&       4019.289			&       0.582           &       $-3.232$\\
\ion{Co}{I}			&       4019.299			&       0.629           &       $-3.769$\\
\element[][13]{CH}	&       4019.315			&       0.463           &       $-6.904$\\
\element[][12]{CH}	&       4019.440			&       1.172           &       $-7.971$\\
\hline
\end{tabular}
\end{center}
\label{tab:ThII4019linelist}
\end{table}

Accurate measurements of thorium abundances, the only element beyond the third r-process peak that we clearly detected in both stars, are rather challenging. The strongest line at $4019.129$\,{\AA} can be significantly blended with a \element[][13]{CH} B-X feature (which corresponds to the clearly visible \element[][12]{CH} B-X feature around $4020$\,{\AA}), along with weak contributions from various other elements (see Fig. \ref{Fig:ThoriumSynthesis} and Table~\ref{tab:ThII4019linelist}). Hence, a very careful synthesis taking into account such blends is required. However, the very low \element[][13]{C} abundance in both stars, as discussed in Sect. \ref{sec:CNO}, simplifies the measurement. We detect a second thorium line at $4250$\,{\AA} in the case of \object{HE~1219$-$0312}, producing an abundance of $\log\epsilon(\rm{Th})=-1.38$, which is $0.09$\,dex lower than the result for the $4019.129$\,{\AA} feature. We discarded the line from our analysis, because it was not included in the laboratory measurements of \citet{Nilssonetal:2002b}. Systematic differences between the abundances derived from the two Th lines may therefore arise. However, the good agreement between the two abundances confirms the overall reliability of our measurements. While \object{CS~29491$-$069} exhibits the expected thorium depletion with respect to the Sun, owing to its presumed old age, \object{HE~1219$-$0312} does not appear to be as depleted with severe consequences for its age determination (see Sect. \ref{sec:ages}).

Similar to the case of lead, the quality of our data was not sufficient for a detection of the \ion{U}{II} $3859.57$\,{\AA} line in either star. A reliable upper limit could again not be measured due to the noise level of the spectral data.

\subsection{Error budget}\label{sec:ErrorBudget}

We performed a detailed error analysis on \object{HE~1219$-$0312}, in order to estimate the different uncertainties which accompany abundance measurements. The various contributions were handled as being completely independent. Although approximate, this approach is good enough to provide an idea of the reliability of our analysis.

The manifold approximations in the treatment of the input physics of hydrostatic LTE model atmospheres are difficult to assess, and are thus omitted in the error budget. However, recent results that were obtained from 3D hydrodynamical model atmospheres, as well as NLTE line formation calculations, indicate that their impact on the accuracy of \emph{absolute} abundances may be large.

Equivalent-width analyses and spectrum syntheses were repeated using models with varying stellar parameters. The response to changes of $T_{\mathrm{eff}}$ was expectedly large for all elements. The same held for variations of $\log g$, apart from the known insensitivity of lines of neutral species to the resulting pressure stratification. However, we stress that the \textit{relative} abundances of the lanthanides and actinides have little sensitivity to both temperature and gas pressure, due to their similar electronic structure, as long as practically unblended lines are used for the measurements. The r-process pattern is therefore not significantly affected by the uncertainties of the stellar parameters, and their contribution to the error bars of abundance ratios of such elements was neglected.

The sensitivity of the abundances to small changes of the metallicity in the model atmosphere was found to be very weak for virtually all species. Since mostly weak lines were chosen for the analysis, the influence of the microturbulence parameter $\xi$ was small for most heavy elements, whereas some of the much more abundant light elements suffer more from its uncertainty because of line saturation.

The rather high noise level of the spectra motivated an investigation of the errors induced by the line fits. $\chi^2$ fits of spectral lines and automated continuum fits were used for better measurement accuracy. Additional spectrum syntheses were carried out with varying continuum placement and line core fits, matching the ends of the $1\sigma$ error bars instead of the spectrum. This yields a rather conservative estimate for the fit uncertainties. For line measurements obtained by equivalent-width analysis, a typical fit error was assumed.

Further uncertainties induced by unresolved blends, by line data, such as $\log gf$ values and excitation potentials, as well as by model imperfections, such as the treatment of continuum opacity or the above mentioned input physics for convection, lead to considerable scatter in the line-to-line abundances of each species. The results were therefore averaged, but it is clear that this is only an approximation, since this scatter is not purely statistical.

The contributions to the total uncertainty, $\sigma_{\mathrm{tot}}$, were then combined as the sum of squares, assuming their complete independence. Since the line-to-line scatter, $\sigma_{\mathrm{\log\epsilon}}$, contains the fit uncertainties, $\sigma_{\mathrm{fit}}$, the maximum of both enters the sum to obtain a conservative estimate:
\begin{displaymath}
\sigma^2_{\mathrm{tot}}=\sigma^2_{\mathrm{sys}}+\left[\mathrm{max}\left(\sigma_{\mathrm{fit}},\sigma_{\mathrm{\log\epsilon}}\right)\right]^2.
\end{displaymath}

\subsection{The origin of the heavy elements in a high-entropy-wind scenario}\label{sec:NcapHEW}

Recent dynamical network calculations investigate the properties of an r-process which is embedded in a model of a SN II with an adiabatically expanding high-entropy wind \citep[HEW,][]{Freiburghausetal:1999,Farouqi:2005,Farouqietal:2005,Farouqietal:2008a}. In this scenario, the expanding matter behind the shock front recombines into $\alpha$-particles and heavy ``seed'' nuclei, accompanied by free neutrons. Freezeout fixes their relative abundance fractions as a function of entropy, $S$, which varies between different mass zones in the model; for details see \citet{Farouqietal:2008a} and references 1-8 therein.

HEW zones of the lowest entropy range ($S<50$) then undergo pure charged-particle capture (alpha-process), producing stable and near-stable isotopes in the iron-group region. Zones of the next higher entropy range ($50<S<100$) are still dominated by charged-particle reactions, however already producing quite neutron-rich, so-called ``beta-delayed neutron precursor" isotopes, providing the first neutrons for a primary, very low-density neutron-(re-)capture process in the $80<A<100$ nuclear mass region. Only in the subsequent higher entropy zones, successively increasing ratios of ``free" neutrons to ``seed" nuclei \citep[see e.g. Fig. 1 in][]{Farouqietal:2008a} become available to produce the classical ``weak" ($100<S<150$) and ``main" ($S>150$) neutron-capture r-process components.

Hence, the total nucleosynthetic yield from a HEW scenario appears as an overlay of SN ejecta with multiple components in different entropy ranges. This superposition might explain the occurrence of neutron-capture in terms of a robust ``main" r-process for heavy elements beyond $Z\approx52$ (Te, Xe), accompanied by an alpha-process, which forms the lighter elements in the region between Fe and Mo and which should in principle be uncorrelated with the neutron-capture (r-process) components.

Anti-correlations of the abundances of Sr, Y and Zr, relative to barium and europium, together with the apparent constancy of their respective ratios, have been observed in halo stars with different degrees of heavy-element enrichment (see Sect. \ref{sec:heavyelm}). These findings suggest that an additional LEPP contributes to the production of elements in the mass region of Sr, Y and Zr. Stars that are heavily enriched with europium, such as \object{HE~1219$-$0312}, seem to deviate from the anti-correlation, showing constant [Sr/Eu] and [Y/Eu] over [Eu/Fe], indicating a significant contribution from the r-process itself \citep{Montesetal:2007}.

These observations may be explained in the HEW scenario by an entropy mix that differs between production sites. In order to produce highly r-process rich ejecta, a corresponding mix could be caused either by an incomplete ejection of iron-group elements from the beginning, or a later partial fallback of the outflowing, denser, low-entropy mass zones onto the nascent neutron star. The resulting ``loss'' of lighter alpha-elements Sr to Zr then needs to decline rapidly with increasing entropy: it must vanish in entropy zones which produce barium, europium and heavier elements through a neutron-capture process in order to reproduce the observed robust r-process abundance patterns. In contrary, r-process poor ejecta with high abundances in the Sr to Zr region would be obtained if the ejecta never reached high entropies. The very different abundance patterns of \object{CS~22892$-$052}, an r-process rich star with low Sr to Zr abundances, and \object{HD~122563}, which is r-process poor with high Sr, Y and Zr abundances, point in this direction; these two stars exhibit very different levels of enrichment with lighter and heavier trans-iron elements \citep[Fig. 3 in][]{Farouqietal:2008a}.

Another possible explanation for the observed abundance divergence builds on varying production yields of supernovae, depending on their type \citep[see, e.g.,][]{Quianetal:2008}.

Figure \ref{Fig:abundance_patternsHEW} shows nucleosynthetic yields of a solar-like HEW model, summed over entropy ranges of $10\le S\le280$ and $60\le S\le280$, and compares them to the observed abundances of \object{CS~29491$-$069} and \object{HE~1219$-$0312}. The theoretical yields are scaled to match the respective Gd abundances. The main model parameters, entropy $S$, electron fraction $Y_{\rm{e}}$ and expansion velocity $V_{\rm{exp}}$, are not yet well constrained by current SN II models and therefore assume realistic estimates.

HEW model predictions for strontium, yttrium and zirconium are dominated by alpha processes at low entropies. A contribution from a large entropy range ($10\le S\le280$) seems to over-predict their absolute abundances in both stars, while their ratios are consistent with the observations (see the upper rows in Table \ref{tab:HEWratios}). Assuming an incomplete ejection or fallback scenario, more than 80\,\% of the synthesized Sr, Y and Zr nuclei failed to reach the ISM in both cases. \object{CS~22892$-$052} and \object{CS~31082$-$001} are similarly ``alpha-poor'' \citep[Fig. 3 in][]{Farouqietal:2008a}. This situation is simulated by virtue of introducing a sharp entropy cutoff below $S=60$. The result is an exponential decrease of the Sr and Y abundances while leaving the other elements unchanged.

Most rare-earth elements are well-matched in \object{CS~29491$-$069}, only Eu is not consistent (see Table \ref{tab:HEWratios} and Fig. \ref{Fig:abundance_patternsHEW}). The ``loss'' of enriched material must therefore have already ceased in mass zones of rather low entropies, requiring an exponential decline. The observed europium ``under-abundance'' cannot be explained by the model. While palladium was not measurable in \object{CS~29491$-$069}, \object{HE~1219$-$0312} exhibits a low Pd abundance with respect to the HEW yields, requiring not only incomplete ejection or fallback of alpha-capture material, but also the ``weak'' r-process component as an additional constraint: almost 80\,\% of the HEW-predicted palladium was not ejected into the ISM. Table \ref{tab:HEWratios} quantifies the large discrepancy with the full entropy range model. It is important to note, however, that the palladium abundance was derived using only one \ion{Pd}{I} feature in \object{HE~1219$-$0312}, rendering the abundance ratios more sensitive to uncertainties in the model atmosphere and in the line data.

All main r-process material, produced at high entropies, is fully observed in both stars. The Th abundance yield is very sensitive to the upper entropy limit and to the electron fraction $Y_{e}$, and is therefore not well constrained in the model. The chosen parameter set produces a Th/Eu ratio of $\log\left(\mbox{Th/Eu}\right)_0 = -0.32$, close to the predictions of \citet{Snedenetal:2003}, $\log\left(\mbox{Th/Eu}\right)_0 = -0.35$, and \citet{Kratzetal:2007}, $\log\left(\mbox{Th/Eu}\right)_0 = -0.28$.

\subsection{Age determinations and the reliability of the Th/Eu chronometer}\label{sec:ages}

\begin{table}[tdp]
\caption{Abundance ratios (non-logarithmic) of alpha-capture, weak and main r-process elements as predicted by the HEW model for $10\le S\le280$, compared to observations in \object{CS~29491$-$069} and \object{HE~1219$-$0312}.}
\begin{center}
\begin{tabular}{lrrr}\hline\hline
 Pair          &	HEW		&	\object{CS~29491$-$069}	&	\object{HE~1219$-$0312}\\
\hline
Sr/Y		&	6.81		&	$6.31\pm3.40$				&	$6.17\pm3.33$\\
Sr/Zr		&	1.25		&	$1.10\pm0.63$				&	$1.07\pm0.56$\\
Y/Zr		&	0.18		&	$0.17\pm0.09$				&	$0.17\pm0.08$\\
\hline
Pd/Ba	&	3.99		&							&	$0.78\pm0.47$\\
Pd/Ce	&	8.76		&							&	$1.86\pm1.34$\\
Pd/Nd	&	4.94		&							&	$1.45\pm0.86$\\
Pd/Eu	&	17.67	&							&	$6.46\pm3.73$\\
Pd/Gd	&	6.88		&							&	$1.45\pm0.91$\\
Pd/Dy	&	7.47		&							&	$1.23\pm0.75$\\
Pd/Er	&	12.08	&							&	$2.04\pm1.27$\\

\hline
\end{tabular}
\end{center}
\label{tab:HEWratios}
\end{table}

\begin{table*}[htdp]
\caption{Logarithmic production ratios $\log\left(\mbox{Th/x}\right)_0$ and corresponding decay ages in \object{CS~29491$-$069} and \object{HE~1219$-$0312} in Gyr.}
\begin{center}
\begin{tabular}{lrrrrrrrr}\hline\hline
Pair	&	\multicolumn{2}{c}{Logarithmic production ratios}	&	\multicolumn{3}{c}{\object{CS~29491$-$069}}	&	\multicolumn{3}{c}{\object{HE~1219$-$0312}}\\
		&	Solar residuals	&	HEW $10\le S\le280$	&	Residual age$^{1}$ &	HEW age		& Error$^{2,3}$ &	Residual age$^{1}$ &	HEW age & Error$^{2,3}$\\
\hline
Th/Ba	&  	$-1.39$		& $-0.96$      			& $ 1.9$			& $17.1$    		& $ 9.9$ (3.3) 		& $-6.5$				& $8.7$		& $5.9$ (4.2)\\ 
Th/La	&  	$-0.65$		& $-0.22$      			& $ 0.9$			& $16.5$    		& $ 9.6$ (2.3) 		& $-5.7$				& $9.9$		& $5.0$ (2.8)\\
Th/Ce	&  	$-0.88$		& $-0.62$      			& $17.1$			& $24.6$    		& $10.9$ (5.6) 		& $-0.6$				& $6.8$		& $9.4$ (8.4)\\
Th/Pr	&  	$-0.36$		& $-0.20$      			& $10.3$			& $13.2$    		& $11.7$ (7.0) 		& $-6.5$				& $-3.6$		& $5.6$ (3.7)\\
Th/Nd	&  	$-1.03$		& $-0.87$      			& $10.5$			& $13.4$    		& $10.2$ (4.2) 		& $-2.6$				& $ 0.4$		& $4.8$ (2.3)\\
Th/Sm	&  	$-0.80$		& $-0.71$      			& $12.0$			& $11.8$    		& $11.1$ (6.1) 		& $-0.1$				& $-0.3$		& $6.6$ (5.1)\\
Th/Eu	&  	$-0.43$		& $-0.32$      			& $ 3.0$			& $ 3.8$    		& $ 9.4$ (1.4) 		& $-4.9$				& $-4.1$		& $4.6$ (1.9)\\
Th/Gd	&  	$-0.99$		& $-0.73$      			& $13.5$			& $21.1$    		& $ 9.6$ (2.3) 		& $-0.5$				& $7.1$		& $6.3$ (4.7)\\
Th/Dy	&  	$-1.01$		& $-0.69$      			& $14.0$			& $24.2$    		& $ 9.9$ (3.3) 		& $ 1.8$				& $12.0$		& $5.3$ (3.3)\\
Th/Ho	&	$-0.41$		& $ 0.04$      			& $ 4.4$			& $21.2$    		& $12.6$ (8.4) 		& $-7.3$				& $9.5$		& $7.4$ (6.1)\\
Th/Er	&  	$-0.79$		& $-0.48$      			& $16.8$			& $26.4$    		& $ 9.9$ (3.3) 		& $ 1.8$				& $11.5$		& $5.9$ (4.2)\\
Th/Tm	&  	$ 0.12$		& $ 0.22$      			&        			&           	    		&			& $ 0.0$				& $0.1$		& $7.4$ (6.1)\\
Th/Hf	&  	$-0.46$		& $ 0.20$      			&        			&           			&    			& $-2.1$ 				& $[24.2]$	& $9.4$ (8.4)\\
Th/Os   	&  	$-0.78$		& $-0.93$      			& $[<36.5]$		& $[<24.6]$    		&			&					&			&\\
\hline
\multicolumn{3}{l}{Arithmetic mean} 					& $9.5\pm6.0$		& $17.6\pm6.8$	& 			& $-2.6\pm3.3$			& $4.8\pm5.9$	&\\
\hline
\multicolumn{9}{l}{1: assuming a solar age of 4.6\,Gyr; 2: without $\sigma_{\mathrm{sys}}$, see Sect.~\ref{sec:ErrorBudget}; 3: values in parentheses without the Th abundance uncertainty}
\end{tabular}
\end{center}
\label{tab:Ages}
\end{table*}

Stellar age estimates can be obtained by comparing abundance ratios of thorium, a radioactive element formed by the r-process, and the observed stable rare earth elements. Among these, europium, which is dominantly produced by the r-process, has been successfully used for age dating in some cases in the past \citep[e.g.,][]{Snedenetal:2003}. The derived ages were consistent with the age of the Universe, as determined from CMB observations, and our understanding of the evolution of the Galaxy. However, the Th/Eu chronometer pair has failed for other objects, e.g., \object{CS~31082$-$001}, which exhibits high abundances of thorium and uranium with respect to the lanthanides. This phenomenon was named ``actinide boost'' in the literature and is currently not understood. Our measurements of thorium allow us to investigate an occurrence of this phenomenon in both stars.

The accuracy of decay age estimates is limited by the accuracy with which initial production ratios can be predicted from calculations, apart from the inevitable abundance measurement uncertainties. An inherent difficulty for all current r-process models is that their cosmic site is still unknown; their physical parameters therefore cannot be tightly constrained yet. Together with uncertainties in the nuclear data, this leads to scatter in the theoretical abundance ratios derived from different models. Moreover, production yields for different pairs vary in their sensitivity to model parameters \citep[see, e.g., Fig. 7 in][]{Wanajoetal:2002}. The HEW models in principle allow such age determinations. However, it is important to stress the dependence of the predicted Th abundance on the chosen model parameters (see Sect. \ref{sec:NcapHEW}), which affects the accuracy of the absolute age.

All known highly r-process enhanced stars exhibit a remarkably robust abundance pattern between $Z\geq56$ and $Z<83$, which at the same time largely agrees with solar r-process residuals \citep{Montesetal:2007}. A second age estimate for each star can therefore be derived from the solar residuals of \citet{Arlandinietal:1999}, combined with the solar abundances of \citet{Asplund:2005}, allowing a comparison with the HEW ages. The observed thorium abundance in the Sun was used as a zero point; all residual ages were consequently corrected with the solar age of approximately $4.6$\,Gyr.

The calculations using r-process residuals in \object{CS~29491$-$069} give an average age of $9.5$\,Gyr with a standard deviation of $6.0$\,Gyr for the individual results. The HEW model yields an average of $17.6$\,Gyr with a similarly large scatter of $6.8$\,Gyr. The individual estimates are roughly consistent with typical ages of 11 to 12\,Gyr found for metal-poor halo stars in the past. Table \ref{tab:Ages} shows initial abundance ratios and derived ages for each element pair. In agreement with the abundance pattern presented in Fig. \ref{Fig:abundance_patterns}, it seems unlikely that \object{CS~29491$-$069} is strongly thorium rich and therefore an ``actinide boost'' star.

The case is different for \object{HE~1219$-$0312}, where almost all abundance pairs yield negative ages when compared to the r-process residuals. We determine an average age of $-2.6$\,Gyr with a standard deviation of 3.3\,Gyr. The HEW yields predict an age of about $4.8$\,Gyr with a scatter of $5.9$\,Gyr. The estimate for hafnium is bracketed for the HEW model due to problems with the nuclear data, rendering the synthetic yield unreliable. It is clear that the high Th abundance causes this shift towards low or even negative ages, and the significantly different results for the two stars, which were obtained using the same initial abundance ratios. We find $\log\left(\mbox{Th/Eu}\right) = -0.23\pm 0.10$ in \object{HE~1219$-$0312}, which is almost identical to the value observed in \object{CS~31082$-$001} ($\log\left(\mbox{Th/Eu}\right)=-0.22\pm 0.12$; \citealt{Hilletal:2002}). Its stellar matter could therefore have experienced an ``actinide boost''. In order to find further proof whether this ``boost" exists or not and if it only applies to actinides, it would be necessary to confirm an expected overabundance of third-peak-elements. Unfortunately, none of these were detectable in \object{HE~1219$-$0312}. The lead abundance is considered an important test for an ``actinide boost'' scenario, as it lies in the decay paths of thorium and uranium. However, \citet{Plezetal:2004} found a low Pb abundance for \object{CS~31082$-$001}, which is hard to reconcile with the expectations from nuclear physics. The nature of the high actinide abundances in this star remains unclear.

An interesting feature in the distribution of individual age estimates derived from r-process residuals is an apparent trend towards higher ages with increasing atomic number of the stable partner element, which is not easily visible in the abundance plots of Fig. \ref{Fig:abundance_patterns}. The thorium abundance uncertainty does not affect the relative age scatter and was therefore removed from the error bars (written in parentheses in Table \ref{tab:Ages}). While both stars seem to exhibit this behavior, \object{CS~29491$-$069} has stronger variation between low and high ages, where \object{HE~1219$-$0312} has a more uniform distribution. The HEW ages follow a weaker gradient than the residuals. The trend might nevertheless point towards an explanation for an ``actinide boost'' phenomenon through an increasing deficiency in lighter elements, causing thorium to appear over-abundant. A possible mechanism could be found in the above mentioned incomplete ejection or fallback scenarios. It is clear that our results are far from decisive in that respect, due to the uncertainties of the predicted initial ratios and the abundance determinations of the stable partner elements, but they may motivate future research.

\section{Discussion and conclusions}\label{sec:DiscussionConclusions}

Currently there are 12 r-II stars reported in the literature: \object{CS~22892$-$052} \citep{Snedenetal:1996}, \object{CS~31082$-$001} \citep{Cayreletal:2001,Hilletal:2002}, \object{CS~29497$-$004} \citep{HERESpaperI}, \object{CS~22183$-$031} \citep{Hondaetal:2004b}, \object{HE~1523$-$0901} \citep{Frebeletal:2007b}, and seven additional stars published in \citet{HERESpaperII}. Among these stars, published abundance analyses based on high-resolution spectroscopy of sufficient quality to detect the \ion{Th}{II} $4019$\,{\AA} line were previously available for only four stars: \object{CS~22892$-$052}, \object{CS~31082$-$001}, \object{CS~29497$-$004}, and \object{HE~1523$-$0901}. With \object{CS~29491$-$069} and \object{HE~1219$-$0312}, we add two stars to this well-studied sample, for a total of six.

The relative abundances of most neutron-capture elements that we analyzed are consistent with their corresponding solar residuals, besides a slight upward trend with the atomic number. However, the progenitor gas cloud of \object{HE~1219$-$0312} may have experienced a particularly strong enrichment with the heaviest elements. This leads to a failure of the commonly used Th/Eu chronometer, along with most other element pairs, by resulting in a negative radioactive decay age. Selective enhancement of elements in the third r-process peak and beyond requires r-process models with high neutron densities (and entropies in a HEW scenario). The plausibility of such a physical environment therefore needs to be further investigated. Measuring abundances of lead and other third-peak elements, which should exhibit similar enrichment as thorium, could contribute to solving this problem. The low Pb abundance of \object{CS~31082$-$001} found by \citet{Plezetal:2004}, however, seems to point towards a more complicated picture. The slight trend in the individual decay age estimates could indicate an alternative explanation of strong actinide enrichment by a deficiency of lighter elements, but the scatter is too large to provide adequate evidence.

Note that \citet{Hondaetal:2004b} reported an increased Th/Eu ratio of $\log\left(\mathrm{Th/Eu}\right)=-0.10$ for \object{CS~30306$-$132}, an r-I giant star with $\mathrm{[Fe/H]}=-2.42$ and $\mathrm{[Eu/Fe]}=+0.85$. The case appears similar to that of \object{HE~1219$-$0312}, with \object{CS~30306$-$132} having a relative thorium overabundance. Likewise, there are no measurements of other third-peak elements or actinides to further investigate a process that causes strong actinide enrichment.

\object{CS~29491$-$069} seems to have a significantly smaller thorium abundance with respect to the overall enrichment with heavy elements; an actinide overabundance is therefore unlikely. Radioactive dating based on solar r-process residuals results in an average age of $9.5$\,Gyr, and $17.6$\,Gyr for the HEW predictions. The Th/Eu pair seems to yield a much younger age, caused by the low europium abundance. The large scatter in decay ages found for different element pairs confirms that stellar chronometry needs to be based on more than one abundance ratio. The accuracy of absolute ages that are determined from theoretical predictions is still limited, owing to uncertainties in the current nucleosynthesis models and to the unknown astrophysical site of the r-process, apart from the inevitable measurement errors.

We also compare our abundance measurements with the yields of recent dynamical network calculations in the framework of a high-entropy-wind (HEW) scenario. Most heavy elements beyond the second r-process peak show good agreement with the predictions. The model matches our observed abundance ratios of strontium, yttrium and zirconium; however, their absolute values are significantly over-predicted. Further supported by the mismatch of palladium in \object{HE~1219$-$0312}, this may be interpreted as an indication of an incomplete ejection or fallback scenario for lighter elements, or as contributions from different types of SN II.

\begin{acknowledgements}
  
We are grateful to the ESO staff at Paranal and Garching for obtaining the observations and reducing the data, respectively. N.C. acknowledges financial support from Deutsche Forschungsgemeinschaft through grants Ch~214/3 and Re~353/44, and by the Knut and Alice Wallenberg Foundation. K.E., A.J.K. and P.S.B acknowledge support of the Swedish Research Council. P.S.B is a Royal Swedish Academy of Sciences Research Fellow supported by a grant from the Knut and Alice Wallenberg Foundation. T.C.B. acknowledges support by the US National Science Foundation under grant AST 07-07776, as well as from grants PHY 02-16783 and PHY 08-22648; Physics Frontier Center/Joint Institute for Nuclear Astrophysics (JINA). K.F., B.P. and K.-L.K. acknowledge financial support by the Deutsche Forschungsgemeinschaft (DFG) under contract KR 806/13-1 and the Helmholtz-Gemeinschaft under grant VH-VI-061.

\end{acknowledgements}

%
\bibliographystyle{aa}
\bibliography{HERESIV}
%
%
\Online
\onecolumn
\tablecaption{Line list used for the abundance analysis of \object{CS~29491$-$069} and \object{HE~1219$-$0312}.}
\label{tab:CS29491Linelist}
\begin{center}
\tablefirsthead{\hline\hline
                                &               &                       &                                       &                       &                               & \multicolumn{3}{c}{\object{CS~29491$-$069}} & \multicolumn{3}{c}{\object{HE~1219$-$0312}}\\
                        Z       &       Atom    &       Ion     &       $\lambda$\,[{\AA}]      &       $\log gf$     &       $\chi$\,[eV]    &       $W_{\lambda}$\,[m{\AA}] &       $\log\epsilon$  &       method    &       $W_{\lambda}$\,[m{\AA}] &       $\log\epsilon$  &       method\\                                              
                        \hline}

                        \tablehead{\hline
                                                        &               &                       &                                       &                       &                               & \multicolumn{3}{c}{\object{CS~29491$-$069}} & \multicolumn{3}{c}{\object{HE~1219$-$0312}}\\
                        Z       &       Atom    &       Ion     &       $\lambda$\,[{\AA}]      &       $\log gf$     &       $\chi$\,[eV]    &       $W_{\lambda}$\,[m{\AA}] &       $\log\epsilon$  &       method    &       $W_{\lambda}$\,[m{\AA}] &       $\log\epsilon$  &       method\\                      
                        \hline}

                        \tabletail{\hline
                        \multicolumn{9}{r}{\small\slshape continued on next page}   \\
                        \hline}

                        \tablelasttail{\hline}

\begin{supertabular}{r l c c r c r l l r l l}
11 & Na  &  1 &   5889.951  & $0.117$  &   0.00 & --       & --       & --     & 105.3  &  3.057  & gauss\\
11 & Na  &  1 &   5895.924  & $-0.184$ &   0.00 & --       & --       & --     &  81.2  &  2.923  & gauss\\
12 & Mg  &  1 &   3336.674  & $-1.230$ &   2.72 & --       & --       & --     &  82.4  &  4.981  & gauss\\
12 & Mg  &  1 &   3829.355  & $-0.231$ &   2.71 &  153.9   &   5.194  & gauss  & --     & --      & --   \\
12 & Mg  &  1 &   3903.859  & $-0.511$ &   4.35 &   34.5   &   5.012  & gauss  & --     & --      & --   \\
12 & Mg  &  1 &   4057.505  & $-1.201$ &   4.35 & --       & --       & --     &  18.6  & 5.253   & gauss\\
12 & Mg  &  1 &   4167.271  & $-1.004$ &   4.35 &   41.0   &   5.620  & gauss  &  19.8  & 5.083   & gauss\\
12 & Mg  &  1 &   4571.096  & $-5.691$ &   0.00 &   30.7   &   5.496  & gauss  & --     & --      & --   \\
12 & Mg  &  1 &   5172.684  & $-0.402$ &   2.71 & --       & --       & --     & 141.1  & 4.754   & gauss\\
12 & Mg  &  1 &   5183.604  & $-0.180$ &   2.72 & --       & --       & --     & 160.7  & 4.780   & gauss\\
12 & Mg  &  1 &   5528.405  & $-0.620$ &   4.35 & --       & --       & --     &  37.1  & 5.011   & gauss\\[0.5ex]

13 & Al  &  1 &   3944.006  & $-0.623$ &   0.00 &  132.1   &   3.872  & gauss  & 103.5  & 3.161   & gauss\\[0.5ex]

14 & Si  &  1 &   3905.523  & $-1.090$ &   1.91 &     --   &   5.20   & synt   & --     & 4.84    & synt \\[0.5ex]

20 & Ca  &  1 &   4226.728  & $ 0.265$ &   0.00 &  170.1   &   3.773  & gauss  & 138.6  & 3.238   & gauss\\
20 & Ca  &  1 &   4283.011  & $-0.292$ &   1.89 &   49.0   &   4.191  & gauss  &  27.6  & 3.600   & gauss\\
20 & Ca  &  1 &   4289.367  & $-0.388$ &   1.88 &   40.3   &   4.119  & gauss  &  20.4  & 3.513   & gauss\\
20 & Ca  &  1 &   4302.528  & $ 0.183$ &   1.90 &   74.0   &   4.201  & gauss  & --     & --      & --   \\
20 & Ca  &  1 &   4318.652  & $-0.295$ &   1.90 &   42.3   &   4.082  & gauss  &  27.0  & 3.601   & gauss\\
20 & Ca  &  1 &   4425.437  & $-0.286$ &   1.88 &   35.0   &   3.906  & gauss  &  19.3  & 3.370   & gauss\\
20 & Ca  &  1 &   4434.957  & $ 0.066$ &   1.89 &   55.1   &   3.934  & gauss  & --     & --      & --   \\
20 & Ca  &  1 &   4435.679  & $-0.412$ &   1.89 &   42.0   &   4.172  & gauss  & --     & --      & --   \\
20 & Ca  &  1 &   4454.779  & $ 0.335$ &   1.90 &   67.2   &   3.904  & gauss  &  50.9  & 3.426   & gauss\\
20 & Ca  &  1 &   4455.887  & $-0.414$ &   1.90 &   28.5   &   3.919  & gauss  & --     & --      & --   \\[0.5ex]

21 & Sc  &  2 &   3535.714  & $-0.465$ &   0.31 &   56.0   &   0.765  & gauss  & --     & --      & --   \\
21 & Sc  &  2 &   3567.696  & $-0.476$ &   0.00 &   74.3   &   0.870  & gauss  & --     & --      & --   \\
21 & Sc  &  2 &   3576.340  & $ 0.007$ &   0.01 &   96.5   &   1.020  & gauss  & --     & --      & --   \\
21 & Sc  &  2 &   4246.822  & $ 0.242$ &   0.31 &   94.2   &   0.731  & gauss  &  85.5  & 0.146   & gauss\\
21 & Sc  &  2 &   4314.083  & $-0.096$ &   0.62 &   69.2   &   0.783  & gauss  &  54.7  & 0.139   & gauss\\
21 & Sc  &  2 &   4400.389  & $-0.536$ &   0.61 &   41.1   &   0.650  & gauss  &  26.5  & 0.011   & gauss\\
21 & Sc  &  2 &   4415.557  & $-0.668$ &   0.60 &   35.7   &   0.667  & gauss  &  24.2  & 0.077   & gauss\\[0.5ex]

22 & Ti  &  1 &   3635.462  & $ 0.047$ &   0.00 &   54.4   &   2.782  & gauss  & --     & --      & --   \\
22 & Ti  &  1 &   3729.807  & $-0.351$ &   0.00 &   40.7   &   2.841  & gauss  & --     & --      & --   \\
22 & Ti  &  1 &   3741.059  & $-0.213$ &   0.02 &   40.9   &   2.723  & gauss  & --     & --      & --   \\
22 & Ti  &  1 &   3904.783  & $ 0.284$ &   0.90 &   15.8   &   2.524  & gauss  & --     & --      & --   \\
22 & Ti  &  1 &   3948.670  & $-0.468$ &   0.00 &   42.7   &   2.934  & gauss  & --     & --      & --   \\
22 & Ti  &  1 &   3958.206  & $-0.177$ &   0.05 &   49.7   &   2.837  & gauss  & --     & --      & --   \\
22 & Ti  &  1 &   3989.759  & $-0.198$ &   0.02 &   46.3   &   2.752  & gauss  &  28.8  & 2.090   & gauss\\
22 & Ti  &  1 &   3998.636  & $-0.056$ &   0.05 &   49.9   &   2.710  & gauss  &  34.4  & 2.096   & gauss\\
22 & Ti  &  1 &   4533.241  & $ 0.476$ &   0.85 &   33.5   &   2.654  & gauss  &  19.2  & 2.053   & gauss\\
22 & Ti  &  1 &   4534.776  & $ 0.280$ &   0.84 &   25.6   &   2.665  & gauss  &  14.3  & 2.080   & gauss\\
22 & Ti  &  1 &   4535.568  & $ 0.162$ &   0.83 &   20.3   &   2.638  & gauss  &  11.0  & 2.056   & gauss\\
22 & Ti  &  1 &   4981.731  & $ 0.504$ &   0.85 &   --     &   --     & --     &  24.0  & 2.114   & gauss\\
22 & Ti  &  1 &   4991.065  & $ 0.380$ &   0.84 &   --     &   --     & --     &  18.7  & 2.085   & gauss\\

22 & Ti  &  2 &   3913.468  & $-0.410$ &   1.12 &   --     &   --     & --     &  77.5  & 2.016   & gauss\\
22 & Ti  &  2 &   4012.385  & $-1.750$ &   0.57 &   72.6   &   2.939  & gauss  & --     & --      & --   \\
22 & Ti  &  2 &   4028.343  & $-0.990$ &   1.89 &   --     &   --     & --     &  20.0  & 2.193   & gauss\\
22 & Ti  &  2 &   4290.219  & $-0.930$ &   1.16 &   --     &   --     & --     &  62.1  & 2.142   & gauss\\
22 & Ti  &  2 &   4394.051  & $-1.770$ &   1.22 &   --     &   --     & --     &  17.6  & 2.107   & gauss\\
22 & Ti  &  2 &   4395.033  & $-0.510$ &   1.08 &   96.0   &   2.731  & gauss  &  79.6  & 2.003   & gauss\\
22 & Ti  &  2 &   4395.850  & $-1.970$ &   1.24 &   --     &   --     & --     &  10.5  & 2.071   & gauss\\
22 & Ti  &  2 &   4399.772  & $-1.220$ &   1.24 &   --     &   --     & --     &  41.9  & 2.114   & gauss\\
22 & Ti  &  2 &   4417.719  & $-1.230$ &   1.16 &   64.7   &   2.804  & gauss  &  46.9  & 2.133   & gauss\\
22 & Ti  &  2 &   4443.794  & $-0.700$ &   1.08 &   87.8   &   2.709  & gauss  &  72.5  & 2.014   & gauss\\
22 & Ti  &  2 &   4444.558  & $-2.210$ &   1.12 &   --     &   --     & --     &  10.8  & 2.176   & gauss\\
22 & Ti  &  2 &   4450.482  & $-1.510$ &   1.08 &   --     &   --     & --     &  33.6  & 2.065   & gauss\\
22 & Ti  &  2 &   4464.450  & $-1.810$ &   1.16 &   --     &   --     & --     &  23.2  & 2.224   & gauss\\
22 & Ti  &  2 &   4468.507  & $-0.600$ &   1.13 &   93.0   &   2.783  & gauss  &  76.6  & 2.061   & gauss\\
22 & Ti  &  2 &   4501.273  & $-0.760$ &   1.12 &   85.1   &   2.731  & gauss  &  73.1  & 2.118   & gauss\\
22 & Ti  &  2 &   4533.969  & $-0.540$ &   1.24 &   86.3   &   2.663  & gauss  &  72.1  & 2.008   & gauss\\
22 & Ti  &  2 &   4563.761  & $-0.790$ &   1.22 &   78.7   &   2.712  & gauss  &  60.9  & 2.003   & gauss\\
22 & Ti  &  2 &   4571.968  & $-0.230$ &   1.57 &   77.7   &   2.505  & gauss  &  62.3  & 1.869   & gauss\\
22 & Ti  &  2 &   4589.958  & $-1.620$ &   1.24 &   --     &   --     & --     &  25.0  & 2.151   & gauss\\[0.5ex]

23 & V   &  1 &   3855.841  & $ 0.013$ &   0.07 &   13.9   &   1.829  & gauss  & --     & --      & --   \\
23 & V   &  1 &   4111.774  & $ 0.408$ &   0.30 &    8.5   &   1.409  & gauss  & --     & --      & --   \\
23 & V   &  1 &   4379.230  & $ 0.580$ &   0.30 &   13.3   &   1.431  & gauss  & --     & --      & --   \\
23 & V   &  1 &   4384.712  & $ 0.510$ &   0.29 &   13.3   &   1.485  & gauss  & --     & --      & --   \\
23 & V   &  1 &   4389.976  & $ 0.200$ &   0.28 &   11.8   &   1.721  & gauss  & --     & --      & --   \\

23 & V   &  2 &   3504.444  & $-0.714$ &   1.10 &   25.8   &   1.498  & gauss  & --     & --      & --   \\
23 & V   &  2 &   3517.296  & $-0.208$ &   1.13 &   44.9   &   1.450  & gauss  & --     & --      & --   \\
23 & V   &  2 &   3530.760  & $-0.470$ &   1.07 &   36.7   &   1.471  & gauss  &  18.8  & 0.709   & gauss\\
23 & V   &  2 &   3545.194  & $-0.259$ &   1.10 &   --     &   --     & --     &  28.4  & 0.773   & gauss\\
23 & V   &  2 &   3592.021  & $-0.263$ &   1.10 &   --     &   --     & --     &  33.8  & 0.891   & gauss\\
23 & V   &  2 &   3589.749  & $-0.295$ &   1.07 &   59.5   &   1.800  & gauss  & --     & --      & --   \\
23 & V   &  2 &   3727.343  & $-0.231$ &   1.69 &   19.7   &   1.449  & gauss  & --     & --      & --   \\
23 & V   &  2 &   3951.960  & $-0.784$ &   1.48 &   16.8   &   1.633  & gauss  & --     & --      & --   \\
23 & V   &  2 &   4005.705  & $-0.522$ &   1.82 &   16.1   &   1.709  & gauss  & --     & --      & --   \\[0.5ex]

24 & Cr  &  1 &   3578.684  & $ 0.409$ &   0.00 &   98.9   &   2.903  & gauss  & --     & --      & --   \\
24 & Cr  &  1 &   3593.481  & $ 0.307$ &   0.00 &   97.2   &   2.955  & gauss  & --     & --      & --   \\
24 & Cr  &  1 &   4254.332  & $-0.114$ &   0.00 &   88.1   &   2.791  & gauss  &  82.9  & 2.334   & gauss\\
24 & Cr  &  1 &   4274.796  & $-0.231$ &   0.00 &   86.7   &   2.865  & gauss  &  74.4  & 2.219   & gauss\\
24 & Cr  &  1 &   4289.716  & $-0.361$ &   0.00 &   79.8   &   2.805  & gauss  &  73.2  & 2.315   & gauss\\
24 & Cr  &  1 &   5206.038  & $ 0.019$ &   0.94 &   --     &   --     & --     &  46.9  & 2.332   & gauss\\
24 & Cr  &  1 &   5345.801  & $-0.980$ &   1.00 &   --     &   --     & --     &  11.8  & 2.552   & gauss\\
24 & Cr  &  1 &   5409.772  & $-0.720$ &   1.03 &   --     &   --     & --     &  16.4  & 2.484   & gauss\\[0.5ex]

25 & Mn  &  1 &   4030.753  & $-0.470$ &   0.00 &   --     &   2.09   & synt HFS  & --     & 1.60    & synt HFS\\
25 & Mn  &  1 &   4033.062  & $-0.618$ &   0.00 &   --     &   2.11   & synt HFS  & --     & 1.58    & synt HFS\\
25 & Mn  &  1 &   4034.483  & $-0.811$ &   0.00 &   --     &   2.14   & synt HFS  & --     & 1.63    & synt HFS\\

25 & Mn  &  2 &   3460.316  & $-0.540$ &   1.81 &   --     &   2.15   & synt HFS  & --     & 1.96    & synt HFS\\
25 & Mn  &  2 &   3488.677  & $-0.860$ &   1.85 &   --     &   2.28   & synt HFS  & --     & 2.05    & synt HFS\\[0.5ex]

26 & Fe  &  1 &   3536.556  & $ 0.115$ &   2.88 &   60.7   &   4.731  & gauss  & --     & --      & --   \\
26 & Fe  &  1 &   3554.925  & $ 0.538$ &   2.83 &   78.9   &   4.746  & gauss  & --     & --      & --   \\
26 & Fe  &  1 &   3565.379  & $-0.133$ &   0.96 &  146.1   &   4.838  & gauss  & --     & --      & --   \\
26 & Fe  &  1 &   3606.679  & $ 0.323$ &   2.69 &   78.1   &   4.878  & gauss  & --     & --      & --   \\
26 & Fe  &  1 &   3651.467  & $ 0.021$ &   2.76 &   56.7   &   4.617  & gauss  & --     & --      & --   \\
26 & Fe  &  1 &   3687.457  & $-0.833$ &   0.86 &  129.0   &   5.176  & gauss  & --     & --      & --   \\
26 & Fe  &  1 &   3694.006  & $ 0.078$ &   3.04 &   76.3   &   5.288  & gauss  & --     & --      & --   \\
26 & Fe  &  1 &   3709.246  & $-0.646$ &   0.92 &  120.9   &   4.904  & gauss  & --     & --      & --   \\
26 & Fe  &  1 &   3758.233  & $-0.027$ &   0.96 &  170.1   &   4.795  & gauss  & --     & --      & --   \\
26 & Fe  &  1 &   3763.789  & $-0.238$ &   0.99 &  152.1   &   4.887  & gauss  & --     & --      & --   \\
26 & Fe  &  1 &   3815.840  & $ 0.237$ &   1.49 &  145.6   &   4.795  & gauss  & --     & --      & --   \\
26 & Fe  &  1 &   3824.444  & $-1.362$ &   0.00 &  144.8   &   5.009  & gauss  & --     & --      & --   \\
26 & Fe  &  1 &   3856.372  & $-1.286$ &   0.05 &  --      &   --     & --     & 133.5  & 4.553   & gauss\\
26 & Fe  &  1 &   3859.911  & $-0.710$ &   0.00 &  191.0   &   4.754  & gauss  & 179.6  & 4.446   & gauss\\
26 & Fe  &  1 &   3865.523  & $-0.982$ &   1.01 &  --      &   --     & --     & 105.3  & 4.607   & gauss\\
26 & Fe  &  1 &   3878.018  & $-0.914$ &   0.96 &  --      &   --     & --     & 109.0  & 4.562   & gauss\\

26 & Fe  &  1 &   3899.707  & $-1.531$ &   0.09 &  122.7   &   4.900  & gauss  & 122.8  & 4.615   & gauss\\
26 & Fe  &  1 &   3920.258  & $-1.746$ &   0.12 &  113.7   &   4.960  & gauss  & --     & --      & --   \\
26 & Fe  &  1 &   3922.912  & $-1.651$ &   0.05 &  121.8   &   4.959  & gauss  & --     & --      & --   \\
26 & Fe  &  1 &   3997.392  & $-0.479$ &   2.73 &   51.1   &   4.812  & gauss  &  49.2  & 4.541   & gauss\\
26 & Fe  &  1 &   4005.242  & $-0.610$ &   1.56 &  100.1   &   4.919  & gauss  & --     & --      & --   \\
26 & Fe  &  1 &   4021.867  & $-0.729$ &   2.76 &   42.4   &   4.906  & gauss  & --     & --      & --   \\
26 & Fe  &  1 &   4032.628  & $-2.377$ &   1.49 &   26.5   &   4.842  & gauss  & --     & --      & --   \\
26 & Fe  &  1 &   4045.812  & $ 0.280$ &   1.49 &  153.4   &   4.790  & gauss  & --     & --      & --   \\
26 & Fe  &  1 &   4062.441  & $-0.862$ &   2.85 &   29.0   &   4.839  & gauss  & --     & --      & --   \\
26 & Fe  &  1 &   4063.594  & $ 0.062$ &   1.56 &  133.3   &   4.838  & gauss  & --     & --      & --   \\
26 & Fe  &  1 &   4067.978  & $-0.472$ &   3.21 &   32.1   &   4.898  & gauss  & --     & --      & --   \\
26 & Fe  &  1 &   4071.738  & $-0.022$ &   1.61 &  121.0   &   4.783  & gauss  & 118.7  & 4.520   & gauss\\
26 & Fe  &  1 &   4076.629  & $-0.529$ &   3.21 &   32.1   &   4.955  & gauss  & --     & --      & --   \\
26 & Fe  &  1 &   4107.488  & $-0.879$ &   2.83 &   39.3   &   5.056  & gauss  & --     & --      & --   \\
26 & Fe  &  1 &   4114.445  & $-1.303$ &   2.83 &   14.9   &   4.875  & gauss  & --     & --      & --   \\
26 & Fe  &  1 &   4132.058  & $-0.675$ &   1.61 &  101.2   &   5.022  & gauss  &  96.3  & 4.635   & gauss\\
26 & Fe  &  1 &   4132.899  & $-1.006$ &   2.85 &   27.8   &   4.948  & gauss  &  22.6  & 4.603   & gauss\\
26 & Fe  &  1 &   4134.678  & $-0.649$ &   2.83 &   42.0   &   4.879  & gauss  & --     & --      & --   \\
26 & Fe  &  1 &   4136.998  & $-0.453$ &   3.41 &   18.6   &   4.759  & gauss  & --     & --      & --   \\
26 & Fe  &  1 &   4143.868  & $-0.511$ &   1.56 &  104.3   &   4.874  & gauss  & 102.0  & 4.559   & gauss\\
26 & Fe  &  1 &   4147.669  & $-2.104$ &   1.49 &   47.4   &   4.995  & gauss  & --     & --      & --   \\
26 & Fe  &  1 &   4153.900  & $-0.321$ &   3.40 &   28.0   &   4.843  & gauss  &  23.0  & 4.532   & gauss\\
26 & Fe  &  1 &   4154.499  & $-0.688$ &   2.83 &   37.1   &   4.810  & gauss  &  30.3  & 4.447   & gauss\\
26 & Fe  &  1 &   4154.806  & $-0.400$ &   3.37 &   29.1   &   4.917  & gauss  &  20.6  & 4.517   & gauss\\
26 & Fe  &  1 &   4156.799  & $-0.809$ &   2.83 &   36.8   &   4.929  & gauss  &  26.9  & 4.493   & gauss\\
26 & Fe  &  1 &   4157.780  & $-0.403$ &   3.42 &   26.2   &   4.906  & gauss  & --     & --      & --   \\
26 & Fe  &  1 &   4174.913  & $-2.969$ &   0.92 &   38.0   &   5.049  & gauss  &  35.2  & 4.677   & gauss\\
26 & Fe  &  1 &   4175.636  & $-0.827$ &   2.85 &   35.5   &   4.933  & gauss  &  35.5  & 4.712   & gauss\\
26 & Fe  &  1 &   4181.755  & $-0.371$ &   2.83 &   53.0   &   4.825  & gauss  & --     & --      & --   \\
26 & Fe  &  1 &   4182.383  & $-1.180$ &   3.02 &   13.5   &   4.892  & gauss  & --     & --      & --   \\
26 & Fe  &  1 &   4184.892  & $-0.869$ &   2.83 &   29.8   &   4.837  & gauss  &  26.3  & 4.537   & gauss\\
26 & Fe  &  1 &   4187.039  & $-0.548$ &   2.45 &   67.6   &   4.882  & gauss  &  60.0  & 4.480   & gauss\\
26 & Fe  &  1 &   4187.795  & $-0.554$ &   2.42 &   68.8   &   4.890  & gauss  &  66.1  & 4.594   & gauss\\
26 & Fe  &  1 &   4191.431  & $-0.666$ &   2.47 &   57.7   &   4.806  & gauss  & --     & --      & --   \\
26 & Fe  &  1 &   4195.329  & $-0.492$ &   3.33 &   29.2   &   4.972  & gauss  & --     & --      & --   \\
26 & Fe  &  1 &   4199.095  & $ 0.155$ &   3.05 &   61.2   &   4.714  & gauss  & --     & --      & --   \\
26 & Fe  &  1 &   4199.983  & $-4.750$ &   0.09 &   10.4   &   5.172  & gauss  & --     & --      & --   \\
26 & Fe  &  1 &   4202.029  & $-0.708$ &   1.49 &   --     &   --     & --     &  97.6  & 4.550   & gauss\\
26 & Fe  &  1 &   4222.213  & $-0.967$ &   2.45 &   47.0   &   4.866  & gauss  &  42.1  & 4.532   & gauss\\
26 & Fe  &  1 &   4227.427  & $ 0.266$ &   3.33 &   56.0   &   4.761  & gauss  &  48.1  & 4.403   & gauss\\
26 & Fe  &  1 &   4233.603  & $-0.604$ &   2.48 &   61.5   &   4.832  & gauss  &  52.7  & 4.414   & gauss\\
26 & Fe  &  1 &   4238.810  & $-0.233$ &   3.40 &   33.0   &   4.857  & gauss  &  26.8  & 4.529   & gauss\\
26 & Fe  &  1 &   4250.119  & $-0.405$ &   2.47 &   69.9   &   4.802  & gauss  &  64.6  & 4.449   & gauss\\
26 & Fe  &  1 &   4260.474  & $ 0.109$ &   2.40 &   91.1   &   4.691  & gauss  &  86.9  & 4.380   & gauss\\
26 & Fe  &  1 &   4271.154  & $-0.349$ &   2.45 &   79.8   &   4.946  & gauss  &  69.4  & 4.476   & gauss\\
26 & Fe  &  1 &   4271.761  & $-0.164$ &   1.49 &  125.5   &   4.833  & gauss  & --     & --      & --   \\
26 & Fe  &  1 &   4282.403  & $-0.779$ &   2.18 &   68.7   &   4.890  & gauss  &  67.3  & 4.588   & gauss\\
26 & Fe  &  1 &   4325.762  & $ 0.006$ &   1.61 &  124.4   &   4.765  & gauss  & 119.9  & 4.456   & gauss\\
26 & Fe  &  1 &   4352.735  & $-1.287$ &   2.22 &   48.1   &   4.969  & gauss  & --     & --      & --   \\
26 & Fe  &  1 &   4375.930  & $-3.031$ &   0.00 &   85.9   &   5.214  & gauss  &  85.6  & 4.822   & gauss\\
26 & Fe  &  1 &   4383.545  & $ 0.200$ &   1.49 &   --     &   --     & --     & 139.3  & 4.428   & gauss\\
26 & Fe  &  1 &   4404.750  & $-0.142$ &   1.56 &  125.9   &   4.854  & gauss  & 119.8  & 4.509   & gauss\\
26 & Fe  &  1 &   4415.123  & $-0.615$ &   1.61 &  103.7   &   4.946  & gauss  &  98.2  & 4.539   & gauss\\
26 & Fe  &  1 &   4430.614  & $-1.659$ &   2.22 &   30.6   &   4.967  & gauss  & --     & --      & --   \\
26 & Fe  &  1 &   4443.194  & $-1.043$ &   2.86 &  --      &   --     & --     &  15.5  & 4.419   & gauss\\
26 & Fe  &  1 &   4447.717  & $-1.342$ &   2.22 &   43.8   &   4.915  & gauss  &  37.5  & 4.539   & gauss\\
26 & Fe  &  1 &   4459.118  & $-1.279$ &   2.18 &   55.7   &   5.036  & gauss  &  47.3  & 4.612   & gauss\\
26 & Fe  &  1 &   4461.653  & $-3.210$ &   0.09 &   76.8   &   5.215  & gauss  & --     & --      & --   \\
26 & Fe  &  1 &   4466.552  & $-0.600$ &   2.83 &   51.8   &   5.000  & gauss  & --     & --      & --   \\
26 & Fe  &  1 &   4489.739  & $-3.966$ &   0.12 &   --     &   --     & --     &  32.6  & 4.676   & gauss\\
26 & Fe  &  1 &   4494.563  & $-1.136$ &   2.20 &   55.7   &   4.913  & gauss  &  51.6  & 4.574   & gauss\\
26 & Fe  &  1 &   4528.614  & $-0.822$ &   2.18 &   72.6   &   4.936  & gauss  & --     & --      & --   \\
26 & Fe  &  1 &   4736.773  & $-0.752$ &   3.21 &   23.6   &   4.930  & gauss  & --     & --      & --   \\
26 & Fe  &  1 &   4871.318  & $-0.363$ &   2.87 &   --     &   --     & --     &  48.9  & 4.461   & gauss\\
26 & Fe  &  1 &   4890.755  & $-0.394$ &   2.88 &   --     &   --     & --     &  50.7  & 4.536   & gauss\\
26 & Fe  &  1 &   4891.492  & $-0.112$ &   2.85 &   72.0   &   4.878  & gauss  &  59.3  & 4.394   & gauss\\
26 & Fe  &  1 &   4903.310  & $-0.926$ &   2.88 &   --     &   --     & --     &  25.5  & 4.569   & gauss\\
26 & Fe  &  1 &   4918.994  & $-0.342$ &   2.87 &   55.2   &   4.781  & gauss  &  47.4  & 4.408   & gauss\\
26 & Fe  &  1 &   4920.503  & $ 0.068$ &   2.83 &   77.3   &   4.787  & gauss  & --     & --      & --   \\
26 & Fe  &  1 &   4938.814  & $-1.077$ &   2.88 &   24.0   &   4.894  & gauss  &  16.6  & 4.470   & gauss\\
26 & Fe  &  1 &   4939.687  & $-3.340$ &   0.86 &   29.0   &   5.095  & gauss  &  22.8  & 4.631   & gauss\\

26 & Fe  &  2 &   3783.347  & $-3.164$ &   2.28 &   21.3   &   5.038  & gauss  & --     & --      & --   \\
26 & Fe  &  2 &   4178.862  & $-2.500$ &   2.58 &   31.6   &   4.892  & gauss  &  26.2  & 4.523   & gauss\\
26 & Fe  &  2 &   4233.172  & $-1.900$ &   2.58 &   58.5   &   4.822  & gauss  &  58.5  & 4.571   & gauss\\
26 & Fe  &  2 &   4303.176  & $-2.560$ &   2.70 &   39.2   &   5.226  & gauss  & --     & --      & --   \\
26 & Fe  &  2 &   4351.769  & $-2.020$ &   2.70 &   62.3   &   5.156  & gauss  & --     & --      & --   \\
26 & Fe  &  2 &   4385.387  & $-2.680$ &   2.78 &   26.2   &   5.139  & gauss  & --     & --      & --   \\
26 & Fe  &  2 &   4416.830  & $-2.410$ &   2.78 &   22.8   &   4.783  & gauss  &  18.3  & 4.421   & gauss\\
26 & Fe  &  2 &   4489.183  & $-2.970$ &   2.83 &   10.0   &   4.957  & gauss  & --     & --      & --   \\
26 & Fe  &  2 &   4491.405  & $-2.700$ &   2.86 &   13.8   &   4.878  & gauss  & --     & --      & --   \\
26 & Fe  &  2 &   4508.288  & $-2.250$ &   2.86 &   30.2   &   4.871  & gauss  & --     & --      & --   \\
26 & Fe  &  2 &   4515.339  & $-2.450$ &   2.84 &   22.6   &   4.882  & gauss  & --     & --      & --   \\
26 & Fe  &  2 &   4520.224  & $-2.600$ &   2.81 &   20.0   &   4.922  & gauss  &  19.3  & 4.665   & gauss\\
26 & Fe  &  2 &   4522.634  & $-2.030$ &   2.84 &   40.5   &   4.852  & gauss  & --     & --      & --   \\
26 & Fe  &  2 &   4541.524  & $-2.790$ &   2.86 &   11.9   &   4.889  & gauss  &   6.1  & 4.335   & gauss\\
26 & Fe  &  2 &   4549.474  & $-2.020$ &   2.83 &   49.3   &   4.997  & gauss  & --     & --      & --   \\
26 & Fe  &  2 &   4555.893  & $-2.160$ &   2.83 &   29.4   &   4.733  & gauss  &  24.8  & 4.389   & gauss\\
26 & Fe  &  2 &   4923.927  & $-1.320$ &   2.89 &   --     &   --     & --     &  70.1  & 4.513   & gauss\\[0.5ex]

27 & Co  &  1 &   3569.370  & $ 0.370$ &   0.92 &   73.6   &   2.656  & gauss  & --     & --      & --   \\
27 & Co  &  1 &   3845.461  & $ 0.010$ &   0.92 &   62.1   &   2.532  & gauss  &  55.4  & 2.046   & gauss\\
27 & Co  &  1 &   3894.073  & $ 0.100$ &   1.05 &   74.0   &   2.883  & gauss  & --     & --      & --   \\
27 & Co  &  1 &   3995.302  & $-0.220$ &   0.92 &   56.8   &   2.595  & gauss  &  53.9  & 2.207   & gauss\\
27 & Co  &  1 &   4118.767  & $-0.490$ &   1.05 &   42.1   &   2.670  & gauss  &  30.9  & 2.126   & gauss\\
27 & Co  &  1 &   4121.311  & $-0.320$ &   0.92 &   60.0   &   2.750  & gauss  &  46.2  & 2.127   & gauss\\[0.5ex]

28 & Ni  &  1 &   3500.846  & $-1.279$ &   0.17 &   91.5   &   3.936  & gauss  & --     & --      & --   \\
28 & Ni  &  1 &   3515.049  & $-0.211$ &   0.11 &  127.7   &   3.547  & gauss  & --     & --      & --   \\
28 & Ni  &  1 &   3524.535  & $ 0.008$ &   0.03 &  154.4   &   3.524  & gauss  & --     & --      & --   \\
28 & Ni  &  1 &   3566.366  & $-0.236$ &   0.42 &  107.3   &   3.519  & gauss  & --     & --      & --   \\
28 & Ni  &  1 &   3619.386  & $ 0.035$ &   0.42 &  117.8   &   3.414  & gauss  & --     & --      & --   \\
28 & Ni  &  1 &   3775.565  & $-1.393$ &   0.42 &   83.4   &   3.881  & gauss  & --     & --      & --   \\
28 & Ni  &  1 &   3783.524  & $-1.310$ &   0.42 &  --      &   --     & --     &  76.9  & 3.268   & gauss\\
28 & Ni  &  1 &   3807.138  & $-1.205$ &   0.42 &   86.7   &   3.770  & gauss  &  83.8  & 3.348   & gauss\\
28 & Ni  &  1 &   3858.292  & $-0.936$ &   0.42 &   93.0   &   3.649  & gauss  &  89.5  & 3.221   & gauss\\
28 & Ni  &  1 &   5476.900  & $-0.890$ &   1.83 &  --      &   --     & --     &  35.5  & 3.225   & gauss\\[0.5ex]

30 & Zn  &  1 &   4722.153  & $-0.338$ &   4.03 &    8.1   &   2.196  & gauss  & --     & --      & --   \\
30 & Zn  &  1 &   4810.528  & $-0.137$ &   4.08 &   12.8   &   2.266  & gauss  &   5.9  & 1.696   & gauss\\[0.5ex]

38 & Sr  &  2 &   4077.719  & $ 0.170$ &   0.00 &  126.4   &   0.512  & gauss  & 130.0  & 0.286   & gauss\\
38 & Sr  &  2 &   4215.519  & $-0.170$ &   0.00 &  115.9   &   0.611  & gauss  & 118.3  & 0.340   & gauss\\[0.5ex]

39 & Y   &  2 &   3549.005  & $-0.280$ &   0.13 &   31.5   & $-0.291$ & gauss  &  41.9  & $-0.427$& gauss\\
39 & Y   &  2 &   3584.518  & $-0.410$ &   0.10 &   38.4   & $-0.035$ & gauss  & --     & --      & --   \\
39 & Y   &  2 &   3600.741  & $ 0.280$ &   0.18 &   49.4   & $-0.379$ & gauss  &  63.3  & $-0.370$& gauss\\
39 & Y   &  2 &   3601.919  & $-0.180$ &   0.10 &   34.7   & $-0.354$ & gauss  & --     & --      & --   \\
39 & Y   &  2 &   3611.044  & $ 0.110$ &   0.13 &   45.0   & $-0.376$ & gauss  &  49.2  & $-0.655$& gauss\\
39 & Y   &  2 &   3628.705  & $-0.710$ &   0.13 &   17.6   & $-0.244$ & gauss  & --     & --      & --   \\
39 & Y   &  2 &   3633.122  & $-0.100$ &   0.00 &   52.6   & $-0.122$ & gauss  &  52.7  & $-0.511$& gauss\\
39 & Y   &  2 &   3788.694  & $-0.070$ &   0.10 &   48.5   & $-0.222$ & gauss  &  58.3  & $-0.365$& gauss\\
39 & Y   &  2 &   3818.341  & $-0.980$ &   0.13 &   15.7   & $-0.093$ & gauss  &  18.8  & $-0.368$& gauss\\
39 & Y   &  2 &   3950.352  & $-0.490$ &   0.10 &   29.6   & $-0.258$ & gauss  &  35.9  & $-0.495$& gauss\\
39 & Y   &  2 &   3982.594  & $-0.490$ &   0.13 &   30.6   & $-0.212$ & gauss  &  36.2  & $-0.464$& gauss\\
39 & Y   &  2 &   4358.728  & $-1.320$ &   0.10 &   --     & --       & --     &  14.1  & $-0.283$& gauss\\
39 & Y   &  2 &   4374.935  & $ 0.160$ &   0.41 &   47.0   & $-0.262$ & gauss  & --     & --      & --   \\
39 & Y   &  2 &   4398.013  & $-1.000$ &   0.13 &   14.5   & $-0.191$ & gauss  &  17.2  & $-0.473$& gauss\\
39 & Y   &  2 &   4883.680  & $ 0.070$ &   1.08 &   12.1   & $-0.349$ & gauss  &  15.5  & $-0.549$& gauss\\
39 & Y   &  2 &   5087.416  & $-0.170$ &   1.08 &   --     & --       & --     &   9.6  & $-0.566$& gauss\\
39 & Y   &  2 &   5205.724  & $-0.340$ &   1.03 &   --     & --       & --     &   6.4  & $-0.655$& gauss\\[0.5ex]

40 & Zr  &  2 &   3479.029  & $-0.690$ &   0.53 &   --     & --       & --     &  13.1  & 0.130   & gauss\\
40 & Zr  &  2 &   3479.383  & $ 0.170$ &   0.71 &   --     & --       & --     &  42.8  & 0.284   & gauss\\
40 & Zr  &  2 &   3505.682  & $-0.360$ &   0.16 &   37.2   & $ 0.425$ & gauss  &  50.5  & 0.371   & gauss\\
40 & Zr  &  2 &   3551.939  & $-0.310$ &   0.09 &   43.0   & $ 0.428$ & gauss  & --     & --      & --   \\
40 & Zr  &  2 &   3613.102  & $-0.465$ &   0.04 &   49.6   & $ 0.672$ & gauss  & --     & --      & --   \\
40 & Zr  &  2 &   4149.217  & $-0.030$ &   0.80 &   26.6   & $ 0.405$ & gauss  & --     & --      & --   \\
40 & Zr  &  2 &   4156.276  & $-0.776$ &   0.71 &   12.7   & $ 0.636$ & gauss  & --     & --      & --   \\
40 & Zr  &  2 &   4161.213  & $-0.720$ &   0.71 &   14.8   & $ 0.661$ & gauss  &  16.6  & 0.377   & gauss\\
40 & Zr  &  2 &   4208.977  & $-0.460$ &   0.71 &   14.1   & $ 0.371$ & gauss  &  20.8  & 0.236   & gauss\\
40 & Zr  &  2 &   4496.980  & $-0.860$ &   0.71 &   10.3   & $ 0.583$ & gauss  &  11.3  & 0.282   & gauss\\[0.5ex]

46 & Pd  &  1 &   3404.579  & $ 0.320$ &   0.81 &   --     & --       & --     &  17.3  & $-0.246$& gauss\\

56 & Ba  &  2 &   4130.645  & $ 0.680$ &   2.72 &     --   & $-0.10$  & synt HFS& --    & $-0.16$ & synt HFS\\
56 & Ba  &  2 &   4554.029  & $ 0.170$ &   0.00 &     --   & $-0.05$  & synt HFS& --    & --      & --   \\
56 & Ba  &  2 &   4934.076  & $-0.150$ &   0.00 &     --   & $-0.05$  & synt HFS& --    & --      & --   \\
56 & Ba  &  2 &   5853.668  & $-1.000$ &   0.60 &     --   & --       & --      & --    & $-0.12$ & synt HFS\\
56 & Ba  &  2 &   6141.713  & $-0.076$ &   0.70 &     --   & $-0.20$  & synt HFS& --    & --      & --   \\[0.5ex]

57 & La  &  2 &   3849.006  & $-0.450$ &   0.00 &   --     & --       & --      & --    & $-0.87$ & synt HFS\\
57 & La  &  2 &   3949.102  & $ 0.490$ &   0.40 &   --     & --       & --      & --    & $-0.90$ & synt HFS\\
57 & La  &  2 &   4086.709  & $-0.070$ &   0.00 &   --     & $-0.82$ & synt HFS & --    & $-0.81$ & synt HFS\\
57 & La  &  2 &   4196.546  & $-0.300$ &   0.32 &   --     & $-0.85$ & synt HFS  & --     & --      & --   \\
57 & La  &  2 &   4333.753  & $-0.060$ &   0.17 &   --     & $-0.90$ & synt HFS & --     & --      & --   \\[0.5ex]

58 & Ce  &  2 &   3577.456  & $ 0.210$ &   0.47 &    --    & --       & --     &   8.1  & $-0.511$& gauss\\
58 & Ce  &  2 &   3655.844  & $ 0.233$ &   0.32 &    --    & --       & --     &   7.3  & $-0.770$& gauss\\
58 & Ce  &  2 &   3942.151  & $-0.180$ &   0.00 &    --    & --       & --     &  18.9  & $-0.300$& gauss\\
58 & Ce  &  2 &   3999.237  & $ 0.232$ &   0.29 &    --    & --       & --     &  10.3  & $-0.700$& gauss\\
58 & Ce  &  2 &   4031.332  & $ 0.080$ &   0.32 &    --    & --       & --     &   6.5  & $-0.745$& gauss\\
58 & Ce  &  2 &   4073.474  & $ 0.230$ &   0.48 &    9.4   & $-0.172$ & gauss  & --     & --      & --   \\
58 & Ce  &  2 &   4118.143  & $ 0.017$ &   0.70 &    --    & --       & --     &   5.3  & $-0.356$& gauss\\
58 & Ce  &  2 &   4120.827  & $-0.130$ &   0.32 &    --    & --       & --     &   6.6  & $-0.539$& gauss\\
58 & Ce  &  2 &   4127.364  & $ 0.106$ &   0.68 &    --    & --       & --     &   7.7  & $-0.284$& gauss\\
58 & Ce  &  2 &   4137.645  & $ 0.246$ &   0.52 &    7.5   & $-0.264$ & gauss  &  14.5  & $-0.303$& gauss\\
58 & Ce  &  2 &   4222.597  & $-0.301$ &   0.12 &    4.8   & $-0.369$ & gauss  &  10.0  & $-0.408$& gauss\\
58 & Ce  &  2 &   4460.207  & $ 0.171$ &   0.48 &   10.2   & $-0.116$ & gauss  & --     & --      & --   \\
58 & Ce  &  2 &   4486.909  & $-0.090$ &   0.29 &    --    & --       & --     &   7.6  & $-0.581$& gauss\\
58 & Ce  &  2 &   4562.359  & $ 0.310$ &   0.48 &    7.5   & $-0.414$ & gauss  &   8.4  & $-0.730$& gauss\\ 
58 & Ce  &  2 &   4628.161  & $ 0.220$ &   0.52 &    6.6   & $-0.348$ & gauss  &   9.0  & $-0.569$& gauss\\[0.5ex]

59 & Pr  &  2 &   4143.112  & $ 0.609$ &   0.37 &     --   & $-0.95$  & synt HFS& --    & $-1.13$ & synt HFS\\
59 & Pr  &  2 &   4222.934  & $ 0.271$ &   0.06 &     --   & --       & --     & --     & $-1.16$ & synt HFS\\
59 & Pr  &  2 &   4408.819  & $ 0.179$ &   0.00 &     --   & --       & --     & --     & $-1.22$ & synt HFS\\[0.5ex]

60 & Nd  &  2 &   3990.097  & $ 0.130$ &   0.47 &     --   & --       & --     &   18.3 & $-0.363$& gauss\\
60 & Nd  &  2 &   4021.327  & $-0.100$ &   0.32 &    8.8   & $-0.311$ & gauss  &   13.1 & $-0.489$& gauss\\
60 & Nd  &  2 &   4061.080  & $ 0.550$ &   0.47 &   20.3   & $-0.364$ & gauss  &   33.8 & $-0.404$& gauss\\
60 & Nd  &  2 &   4069.265  & $-0.570$ &   0.06 &    7.8   & $-0.189$ & gauss  & --     & --      & --   \\
60 & Nd  &  2 &   4109.071  & $-0.160$ &   0.06 &   17.1   & $-0.205$ & gauss  & --     & --      & --   \\
60 & Nd  &  2 &   4109.448  & $ 0.350$ &   0.32 &   23.5   & $-0.248$ & gauss  & --     & --      & --   \\
60 & Nd  &  2 &   4232.374  & $-0.470$ &   0.06 &    8.6   & $-0.261$ & gauss  &   14.4 & $-0.393$& gauss\\
60 & Nd  &  2 &   4358.161  & $-0.160$ &   0.32 &     --   & --       & --     &   15.2 & $-0.392$& gauss\\
60 & Nd  &  2 &   4446.384  & $-0.350$ &   0.20 &    6.0   & $-0.420$ & gauss  &   12.5 & $-0.447$& gauss\\      
60 & Nd  &  2 &   4462.979  & $ 0.040$ &   0.56 &   10.2   & $-0.167$ & gauss  &   13.0 & $-0.411$& gauss\\[0.5ex]

62 & Sm  &  2 &   3568.271  & $ 0.284$ &   0.49 &   13.4   & $-0.423$ & gauss  & --     & --      & --   \\
62 & Sm  &  2 &   3609.492  & $ 0.156$ &   0.28 &   19.7   & $-0.315$ & gauss  & --     & --      & --   \\
62 & Sm  &  2 &   3661.352  & $-0.357$ &   0.04 &   --     & --       & --     & 21.9   & $-0.400$& gauss\\
62 & Sm  &  2 &   3760.710  & $-0.403$ &   0.19 &   --     & --       & --     & 11.2   & $-0.591$& gauss\\
62 & Sm  &  2 &   3896.972  & $-0.668$ &   0.04 &   --     & --       & --     &  9.9   & $-0.579$& gauss\\
62 & Sm  &  2 &   4318.927  & $-0.246$ &   0.28 &    8.0   & $-0.508$ & gauss  &  14.3  & $-0.595$& gauss\\
62 & Sm  &  2 &   4424.337  & $ 0.140$ &   0.49 &    8.8   & $-0.630$ & gauss  & --     & --      & --   \\
62 & Sm  &  2 &   4519.630  & $-0.352$ &   0.54 &   --     & --       & --     &  6.2   & $-0.612$& gauss\\
62 & Sm  &  2 &   4642.230  & $-0.520$ &   0.38 &   --     & --       & --     &  5.1   & $-0.735$& gauss\\[0.5ex]

63 & Eu  &  2 &   3819.672  & $ 0.510$ &   0.00 &   --     & $-1.06$  & synt HFS& -- & $-1.07$ & synt HFS\\
63 & Eu  &  2 &   3907.107  & $ 0.170$ &   0.21 &   --     & $-1.02$  & synt HFS& -- & $-1.09$ & synt HFS\\
63 & Eu  &  2 &   4129.725  & $ 0.220$ &   0.00 &   --     & $-1.02$  & synt HFS& -- & $-1.03$ & synt HFS\\
63 & Eu  &  2 &   4205.042  & $ 0.210$ &   0.00 &   --     & $-1.01$  & synt HFS& -- & --      & --      \\[0.5ex]

64 & Gd  &  2 &   3336.184  & $-0.457$ &   0.00 &   --     &  --        & --     &  15.2  & $-0.348$& gauss\\
64 & Gd  &  2 &   3423.924  & $-0.520$ &   0.00 &   --     &  --        & --     &  12.1  & $-0.421$& gauss\\
64 & Gd  &  2 &   3549.359  & $ 0.260$ &   0.24 &   22.0   &  $-0.228$  & gauss  &  33.4  & $-0.313$& gauss\\
64 & Gd  &  2 &   3557.058  & $ 0.210$ &   0.60 &   --     &  --        & --     &  13.4  & $-0.427$& gauss\\
64 & Gd  &  2 &   3646.196  & $ 0.328$ &   0.24 &   20.5   &  $-0.332$  & gauss  & --     & --      & --   \\
64 & Gd  &  2 &   3654.624  & $-0.030$ &   0.08 &   19.2   &  $-0.214$  & gauss  &  22.3  & $-0.510$& gauss\\
64 & Gd  &  2 &   3813.977  & $-0.215$ &   0.00 &   --     &  --        & --     &  32.8  & $-0.209$& gauss\\
64 & Gd  &  2 &   3844.578  & $-0.400$ &   0.14 &   --     &  --        & --     &  11.3  & $-0.495$& gauss\\
64 & Gd  &  2 &   3916.509  & $ 0.103$ &   0.60 &    9.1   &  $-0.238$  & gauss  & --     & --      & --   \\
64 & Gd  &  2 &   4037.893  & $-0.230$ &   0.56 &   --     &  --        & --     &   6.3  & $-0.503$& gauss\\
64 & Gd  &  2 &   4049.855  & $ 0.429$ &   0.99 &    8.3   &  $-0.199$  & gauss  & --     & --      & --   \\
64 & Gd  &  2 &   4085.558  & $-0.070$ &   0.73 &    4.0   &  $-0.330$  & gauss  & --     & --      & --   \\
64 & Gd  &  2 &   4130.366  & $-0.090$ &   0.73 &   --     &  --        & --     &   6.4  & $-0.446$& gauss\\
64 & Gd  &  2 &   4191.075  & $-0.570$ &   0.43 &    3.4   &  $-0.249$  & gauss  &   4.6  & $-0.475$& gauss\\
64 & Gd  &  2 &   4251.731  & $-0.365$ &   0.38 &    6.6   &  $-0.204$  & gauss  & --     & --      & --   \\[0.5ex]

66 & Dy  &  2 &   3407.796  & $ 0.180$ &   0.00 &   --     & --         & --     &  50.6  & $-0.263$& gauss\\
66 & Dy  &  2 &   3454.317  & $-0.140$ &   0.10 &   --     & --         & --     &  28.9  & $-0.406$& gauss\\
66 & Dy  &  2 &   3460.969  & $-0.070$ &   0.00 &   --     & --         & --     &  38.7  & $-0.351$& gauss\\
66 & Dy  &  2 &   3506.815  & $-0.440$ &   0.10 &   --     & --         & --     &  17.2  & $-0.445$& gauss\\
66 & Dy  &  2 &   3531.707  & $ 0.790$ &   0.00 &   --     & --         & --     &  68.4  & $-0.332$& gauss\\
66 & Dy  &  2 &   3536.019  & $ 0.530$ &   0.54 &   26.7   &  $-0.290$  & gauss  &  35.1  & $-0.437$& gauss\\
66 & Dy  &  2 &   3538.519  & $-0.020$ &   0.00 &   29.9   &  $-0.250$  & --     &  40.8  & $-0.366$& gauss\\
66 & Dy  &  2 &   3563.148  & $-0.360$ &   0.10 &   --     & --         & --     &  26.4  & $-0.269$& gauss\\
66 & Dy  &  2 &   3694.810  & $-0.110$ &   0.10 &   23.1   &  $-0.251$  & gauss  &  33.9  & $-0.359$& gauss\\
66 & Dy  &  2 &   3869.864  & $-1.050$ &   0.00 &    6.7   &  $-0.134$  & gauss  & --     & --      & --   \\
66 & Dy  &  2 &   3944.681  & $ 0.100$ &   0.00 &   42.8   &  $-0.179$  & gauss  & --     & --      & --   \\
66 & Dy  &  2 &   3983.651  & $-0.310$ &   0.54 &   --     & --         & --     &  12.1  & $-0.361$& gauss\\
66 & Dy  &  2 &   3996.689  & $-0.260$ &   0.59 &   --     & --         & --     &  12.4  & $-0.340$& gauss\\
66 & Dy  &  2 &   4000.450  & $ 0.060$ &   0.10 &   35.6   &  $-0.200$  & gauss  & --     & --      & --   \\
66 & Dy  &  2 &   4077.966  & $-0.040$ &   0.10 &   30.6   &  $-0.227$  & gauss  &  48.7  & $-0.197$& gauss\\
66 & Dy  &  2 &   4111.343  & $-0.850$ &   0.00 &    9.2   &  $-0.213$  & gauss  & --     & --      & --   \\[0.5ex]

67 & Ho  &  2 &   3398.901  & $ 0.410$ &   0.00 &   --     &  --        & --     & --     & $-1.09$ & synt HFS\\
67 & Ho  &  2 &   3810.738  & $ 0.142$ &   0.00 &   --     &  $-1.02$  & synt HFS  & --     & $-1.17$ & synt HFS\\

68 & Er  &  2 &   3559.894  & $-0.694$ &   0.00 &   10.9   &  $-0.315$  & gauss  & --     & --      & --   \\
68 & Er  &  2 &   3616.566  & $-0.306$ &   0.00 &   23.6   &  $-0.275$  & gauss  &  31.4  & $-0.457$& gauss\\
68 & Er  &  2 &   3786.836  & $-0.520$ &   0.00 &   13.8   &  $-0.423$  & gauss  & --     & --      & --   \\
68 & Er  &  2 &   3896.234  & $-0.118$ &   0.05 &   22.4   &  $-0.513$  & gauss  &  31.8  & $-0.655$& gauss\\
68 & Er  &  2 &   3906.312  & $ 0.122$ &   0.00 &   40.8   &  $-0.373$  & gauss  & --     & --      & --   \\[0.5ex]

69 & Tm &   2 &   3848.020  & $-0.130$ &   0.00 &   --     &  --        & --     & --     & $-1.51$ & synt \\[0.5ex]

72 & Hf  &  2 &   3399.793  & $-0.57$ &   0.00 &   --     &  --        & --     & --     & $-0.97$ & synt \\[0.5ex]

76 & Os  &  1 &   4260.849  & $-1.470$ &   0.00 &   --     & $< 0.03$   & synt upper limit  & --     & --      & --   \\[0.5ex]

90 & Th  &  2 &   4019.129  & $-0.228$ &   0.00 &   --     &  $-1.43$   & synt   & --     & $-1.29$  & synt\\\hline
\end{supertabular}
\end{center}

%

\newpage
\tableofcontents

\listofobjects
\end{document}